\documentclass{emulateapj}
\def\Sref#1{\S\ref{Sec:#1}}
\def\Fref#1{Figure~\ref{Fig:#1}}
\def\Tref#1{Table~\ref{Tab:#1}}

\newcommand{\altm} {\altaffilmark}
\newcommand{\rah}   {^{\mathrm h}}  
\newcommand{\ram}   {^{\mathrm m}}  
\newcommand{\ras}   {^{\mathrm s}}  
\newcommand{\dd}   {^\circ}  
\newcommand{\am}   {^\prime} 
\newcommand{\as}   {^{\prime\prime}} 
\newcommand{\kms}   {km~s$^{-1}$}
\newcommand{\mpy}   {mas~yr$^{-1}$} 
\newcommand{\muas}  {\mbox{$\mu$as}} 

\newcommand{\pmra}  {\mu_{\alpha}}
\newcommand{\pmdec} {\mu_{\delta}}

\shorttitle{Parallaxes and Proper Motions for 14 Pulsars}
\shortauthors{Chatterjee et al.}
\slugcomment{Accepted by the Astrophysical Journal}

\begin{document}
\title{Precision Astrometry with the Very Long Baseline
  Array:\\ Parallaxes and Proper Motions for 14 Pulsars}
\author{
S. Chatterjee\altm{1}, W. F. Brisken\altm{2}, W. H. T. Vlemmings\altm{3}, 
W. M. Goss\altm{2}, T. J. W. Lazio\altm{4}, \\J. M. Cordes\altm{5}, 
S. E. Thorsett\altm{6}, E. B. Fomalont\altm{7}, A. G. Lyne\altm{8},
M. Kramer\altm{8,9}  
}
\begin{abstract}
Astrometry can bring powerful constraints to bear on a variety of
scientific questions about neutron stars, including their origins,
astrophysics, evolution, and environments.  Using phase-referenced
observations at the VLBA, in conjunction with pulsar gating and
in-beam calibration, we have measured the parallaxes and proper
motions for 14 pulsars. The smallest measured parallax in our sample
is $0.13 \pm 0.02$~mas for PSR~B1541+09, which has a most probable
distance of $7.2^{+1.3}_{-1.1}$~kpc.  We detail our methods,
including initial VLA surveys to select candidates and find in-beam
calibrators, VLBA phase-referencing, pulsar gating, calibration, and
data reduction.  The use of the bootstrap method to estimate
astrometric uncertainties in the presence of unmodeled systematic
errors is also described.  Based on our new model-independent
estimates for distance and transverse velocity, we investigate the
kinematics and birth sites of the pulsars and revisit models of the
Galactic electron density distribution.  We find that young pulsars
are moving away from the Galactic plane, as expected, and that age
estimates from kinematics and pulsar spindown are generally in
agreement, with certain notable exceptions.  Given its present
trajectory, the pulsar B2045$-$16 was plausibly born in the open
cluster NGC~6604.  For several high-latitude pulsars, the NE2001
electron density model underestimates the parallax distances by a
factor of two, while in others the estimates agree with or are larger
than the parallax distances, suggesting that the interstellar medium
is irregular on relevant length scales.
The VLBA astrometric results for the recycled pulsar J1713+0747 are
consistent with two independent estimates from pulse timing, enabling
a consistency check between the different reference frames.
\end{abstract}

 
\altaffiltext{1}{Sydney Institute for Astronomy, School of Physics,
  The University of Sydney, NSW 2006, Australia;
  s.chatterjee@physics.usyd.edu.au \\
  Current address: Dept.\ of Astronomy, Cornell University, Ithaca, NY 14853}

\altaffiltext{2}{National Radio Astronomy Observatory, 
  Socorro, NM 87801}  
\altaffiltext{3}{Argelander Institute for Astronomy, University of Bonn,
  Auf dem H{\"u}gel 71, 53121 Bonn, Germany} 
\altaffiltext{4}{Remote Sensing Division, Naval Research Laboratory,
  Washington, DC 20375} 
\altaffiltext{5}{Dept.\ of Astronomy, Cornell University, Ithaca, NY 14853}
\altaffiltext{6}{Dept.\ of Astronomy and Astrophysics, University 
  of California, Santa Cruz, CA 95064}
\altaffiltext{7}{National Radio Astronomy Observatory, 
  Charlottesville, VA 22903}  
\altaffiltext{8}{University of Manchester, 
  Jodrell Bank Centre for Astrophysics, 
  Manchester M13 9PL, UK}
\altaffiltext{9}{MPI f\"ur Radioastronomie, Bonn, Germany}

\section{Background and Goals}

Neutron stars are exotic laboratories for some of the most extreme
physics in the Universe.  Precise and accurate measurements of the
position, proper motion, and parallax of neutron stars (NS) can be
obtained by astrometry, and such measurements can be exploited to
address a variety of scientific questions.

For example, measuring the position of an object in two different
coordinate frames permits the very fundamental operation of tying the
two reference frames together. Since they are compact sources and are
accessible to astrometry with different techniques and at different
wavelengths, NS are particularly well suited to such reference frame
ties.  Specifically, the optical counterpart of the radio pulsar
J0437$-$4715 may provide a frame-tie between the optical reference
frame and the radio-defined International Celestial Reference Frame
\citep[ICRF;][]{mae+98} for the Space Interferometry Mission.

Precise proper motions for NS allow them to be traced back to their
birth sites in massive stellar clusters, and to associations with
runaway stars \citep{hbz00,vcc04,cvb+05}.  For very young objects,
associations with their progenitor supernova remnants can be verified
or refuted, leading to independent age estimates for both the NS and
the supernova remnant itself
\citep[e.g.,][]{gf00,tbg02,mgb+02,klh+03,bgc+06,zbcg08}.  

Combined with estimates for their distances, the proper motions of NS
also lead to velocity estimates.  The high velocity tail of the
distribution implies that large kicks are imparted to proto-neutron
stars during core collapse, and a number of mechanisms have been
proposed for these natal kicks \citep[e.g.,][]{bh96,jm96,al99,sbhf05}.
The proper motions of magnetars \citep{hcb+07,dceh08,kch+08} can
potentially test the role of strong magnetic fields in such kicks,
while precise comparisons of the proper motion and spin axis vectors
of pulsars \citep{jhv+05,nr07,kcga08} constrain the degree of rotational
averaging in hydrodynamic kicks \citep[e.g.][]{lcc01}.

When a parallax measurement is possible, a {\em model independent}
estimate is obtained for the distance and velocity of the NS.  Each
such measurement calibrates global models of the Galactic electron
density \citep{tc93,cl02}, thus improving distance estimates from
pulse dispersion measure for the rest of the radio pulsar population,
as well as probing the distribution of electron density in the local
interstellar medium \citep[e.g.,][]{tbm+99,ccl+01}.

Observed thermal radiation from the NS surface can be used, in
combination with a precise distance, to constrain the size of the
photosphere, the NS radius, and thus the Equation of State of matter
at extreme pressures and densities \citep{yp04,lp04}.  For radio
pulsars, uncertainties in the magnetospheric emission restrict such
an exercise to very young and hot objects \citep[e.g.,
PSR~B0656+14,][]{btgg03}, while isolated NS which are X-ray bright and
radio quiet pose a challenge for optical astrometry \citep[e.g.,
RX~J1856.5$-$3754,][]{kvka02,wl02}.

Astrometry can thus bring powerful constraints to bear on a variety of
scientific questions about neutron stars, including their origins,
astrophysics, evolution, and environments.  The primary obstacle is
the difficulty of such astrometric observations.  
As in most other astrometric applications, multiple position
measurements of a given target over some time span are the basic
observables from which astrometric parameters are derived.  For
individual objects, rather than an ensemble of stars over the entire
sky, such measurements are necessarily relative in nature, but
absolute positions can be inferred from measurements relative to
sources that define the ICRF.  Over time, repeated measurements of the
position $\vec\theta$ allow a proper motion $\vec{\mu}$ to be
derived. For a precise proper motion, the primary consideration is a
long time baseline, limited by the stability of the reference frame
and the variability or motion of the frame-defining sources.  Finally,
with enough astrometric precision, a trigonometric parallax $\pi$ may
also be measurable for NS. The primary consideration for such
measurements is appropriate sampling over the Earth's orbital phase,
not just a long time baseline.

Neutron stars emit over a broad range of frequencies, and astrometric
observations of NS have been conducted at bands from radio ($\sim
10^8$~Hz) to X-rays ($\sim 10^{17}$~Hz).  For example, \citet{kva07}
have used optical observations with the Hubble Space Telescope to
determine a proper motion $\mu = 107.8 \pm 1.2$~mas~yr$^{-1}$ and a
parallax $\pi = 2.8 \pm 0.9$~mas for RX~J0720.4$-$3125.  However, only
a few NS can be observed at optical or infra-red frequencies.  In
X-rays, the resolution that can be achieved with the current
generation of telescopes is a limiting factor, but \citet{wp07} have
used the Chandra X-ray Observatory to measure a proper motion of $165
\pm 25$~mas~yr$^{-1}$ for the RX~J0822$-$4300, in the center of the
Puppis~A supernova remnant.

The majority of known NS are radio pulsars, and pulse timing is
routinely used to refine their positions and proper motions.  A subset
of recycled (``millisecond'') pulsars have rotation rates that are
stable enough to permit sub-milliarcsecond astrometry based on pulse
time of arrival.  As a particularly interesting example,
\citet{vbv+08} use pulse timing at Parkes to infer a precise distance
for the binary pulsar~J0437$-$4715 based on the observed orbital
period derivative $\dot{P_b}$ and the assumption that the intrinsic
orbital period derivative is dominated by the emission of
gravitational radiation.  The measurement leads to an estimate for the
pulsar mass and an upper limit on the variation of the gravitational
constant $G$.
However, most pulsars do not have such stable rotation, particularly
when they are young, and Very Long Baseline interferometry (VLBI)
has usually been utilized to determine their astrometric parameters.
Such efforts have a long history \citep[e.g.,][]{gtwr86}, but have
become much more feasible with the Very Long Baseline Array (VLBA),
which provides full-time, dedicated VLBI capabilities with identical
antennas, allowing good control of systematic errors and leading to
many recent parallax measurements
\citep[e.g.,][]{fgbc99,bbb+00,ccl+01,bbgt02,ccv+04}. 

However, not only are most pulsars faint, but the most interesting
categories, namely the youngest pulsars and the recycled ones, appear
to be disproportionately faint.  Pulsar gating (which accumulates
signal only during the predicted on-pulse periods for pulsars) is now
routinely employed to boost the signal-to-noise ratios (S/N) for VLBA
astrometric observations, using contemporaneous pulse timing to
reliably predict the pulse phase at the correlator.  To make matters
worse, however, it is not just a lack of sensitivity but systematic
errors contributed by the ionosphere and the troposphere which are the
primary impediments to sub-milliarcsecond astrometry.  Progress has
been made, for example, with GPS-based ionospheric calibration schemes
(Walker \& Chatterjee 1999, VLBA Scientific Memo\footnote{See {\tt
    http://www.vlba.nrao.edu/memos/sci/}} 23) and with wideband
ionospheric calibration techniques \citep{bbb+00,bbgt02}. In-beam
calibration \citep{fgbc99}, using a faint source in the same primary
beam as the target source, has also proved to be effective
\citep{ccl+01,cvb+05}.

Here we present a set of results from a large astrometric program to
determine the proper motions and parallaxes of radio pulsars.  The
project involved a large scale initial survey at the Very Large Array
(VLA) to identify target pulsars and detect in-beam calibrators, as
discussed in \Sref{inb}, followed by multi-epoch observations with
the VLBA (\Sref{vlbaobs}). We provide details of the calibration and
data reduction in \Sref{reduce}, and our use of the bootstrap method
to estimate uncertainties is discussed in \Sref{analysis}, as a
template for ongoing and future VLBA astrometry programs.  Some
implications of our results are explored in \Sref{discuss},
specifically for NS birth sites, Galactic electron density models, and
reference frame ties. Finally, our conclusions are summarized in
\Sref{final}.

\section{Pulsar Selection and Calibrator Identification}\label{Sec:inb}

Astrometric measurements with the VLBA are conducted in the
International Celestial Reference Frame \citep[ICRF,][]{mae+98,fma+04}.  
Each pulsar of interest thus requires the identification of suitable
nearby calibrators that tie the observations to the ICRF.

In the northern hemisphere, the density of ICRF sources and secondary
calibration sources with sufficiently accurate positions in the ICRF
\citep{vcs1,vcs2,vcs3} is about 0.1~deg$^{-2}$, implying typical
calibrator-target separations of $\sim$2\arcdeg, although the nearest
calibrators can be as much as 5\arcdeg\ distant in some cases.
Astrometric errors due to differential propagation effects through
Earth's ionosphere and troposphere scale with calibrator separation
\citep[e.g.,][]{ccv+04}, requiring smaller angular separations for
higher astrometric accuracy.  The detection and use of compact sources
considerably closer to the target than those in the existing catalogs
offers a straightforward way to improve astrometric precision.
Unfortunately, only a small fraction of the many sufficiently bright
point-like sources typically detected by the VLA, even in its most
extended (highest resolution) configuration, are compact enough ($<
10$~mas) to serve as VLBA phase calibrators, owing either to intrinsic
source size or to scattering by the intervening interstellar medium
(ISM).

In our previous work, the search for such nearby compact calibrators
has led to the development of ``in-beam calibration''
\citep{fgbc99,ccl+01}.  Here a compact source is identified within the
same field of view (i.e., primary beam) as the target source.  For the
25-m dishes of the VLA and VLBA, this implies an angular separation
less than $\sim$25\arcmin\ between the target and the in-beam source
for observations at the upper end of our band ($\sim 1.7$~GHz).  Such
sources are likely to be weak, requiring the use of a stronger, more
distant phase referencing source for initial calibration at each
epoch, but the position of the target is thereafter referenced to the
in-beam calibrator position.  In-beam calibration has the dual benefit
of allowing the calibrator and target to be observed simultaneously,
thereby increasing the sensitivity for both objects, as well as
reducing the angular extrapolation of the calibration solutions and
eliminating the need to interpolate the calibration in time.

\subsection{Initial Pulsar Field Survey}\label{Sec:survey}

An initial sample of 63 pulsars was chosen from the roughly one
thousand pulsars known at the inception of this project in 2001.
Pulsars were selected based primarily on flux density, declination,
and dispersion measure distance estimate, so that precision astrometry
was plausible using techniques refined in our previous work. A few
pulsars were included since they were of particular scientific
interest (e.g., the recycled pulsar~J1713+0747) even though they were
expected to pose technical challenges; sample completeness was not a
consideration.

The field of view around each of the 63 selected pulsars was observed
with the VLA between 2002 February~7 and~9 (project code AC629). Given
the similar 25-m diameters of the VLA and VLBA dishes, they have well
matched primary beams at the same observation frequencies.  Each field
was observed for $\sim$15 minutes at 1.4~GHz with the VLA in its
highest angular resolution A-configuration ($\sim$1\farcs4 at
1.4~GHz).  A spectral-line observing mode was used, with a 23~MHz
spectral window or bandpass, and the correlator output was dumped
every 5~s (as opposed to the more standard 10~s).  The use of a
spectral-line mode and short correlator dump times were both motivated
by the desire to image the full field of view with minimal distortions
due to time-average and bandwidth smearing \citep{bs99}.

In order to detect potential in-beam calibrators, as well as bright
confusing sources within the primary beam and beyond, a low-resolution
image with a diameter of 1\fdg1 was searched for sources.  Small
facets surrounding each source were then imaged and deconvolved at the
full resolution, employing polyhedral imaging \citep{cp92}.  For each
field, self-calibration was used to improve the image dynamic range,
and if a substantial improvement was obtained in the dynamic range,
the source-finding process was repeated in an effort to identify any
weaker sources that might have been missed in the first iteration.

Each detected source was inspected for compactness and cataloged with
its brightness, position, and angular separation from the pulsar.  A
total of~1060 sources stronger than 1~mJy were detected in the 63
fields; of these, 269 were within 25\arcmin\ of the target pulsar and
were deemed compact enough (unresolved at 1\farcs4 resolution) to
warrant further investigation.  At least one such candidate in-beam
calibrator was found for 62 of the 63 target pulsars, and an example
is shown in \Fref{inb}.
In addition, the position of each pulsar was measured to a precision
of between 0\farcs2 and 0\farcs05, sufficient to enable the later VLBA
observations.

The angular resolution of these preliminary VLA observations was still
a factor of 100 below that of the planned VLBA observations.  A second
set of VLA observations at 8.4~GHz was used to determine the
sub-arcsecond structure and verify the compactness of the candidate
in-beam calibrators, as well as to determine their spectral indices.
Each of the 269 candidate in-beam calibrators was observed in
individual pointings on 2002 April~6 with the VLA A-configuration at
8.4~GHz, which provides a resolution $\sim$0\farcs2.  Each source was
observed for approximately 3~minutes in continuum mode with two 50~MHz
spectral windows.  Sources that were resolved were rejected, leaving
97 compact sources as potential in-beam calibrators with the VLBA,
with at least one for all but eight of the original 63 pulsars.  The
refinement process is illustrated in \Fref{inb}, and details of all
the candidate sources are archived
online\footnote{\tt http://www.astro.cornell.edu/\~{}shami/psrvlb/}.

\begin{figure*}[t!]
\epsscale{0.85}
\plotone{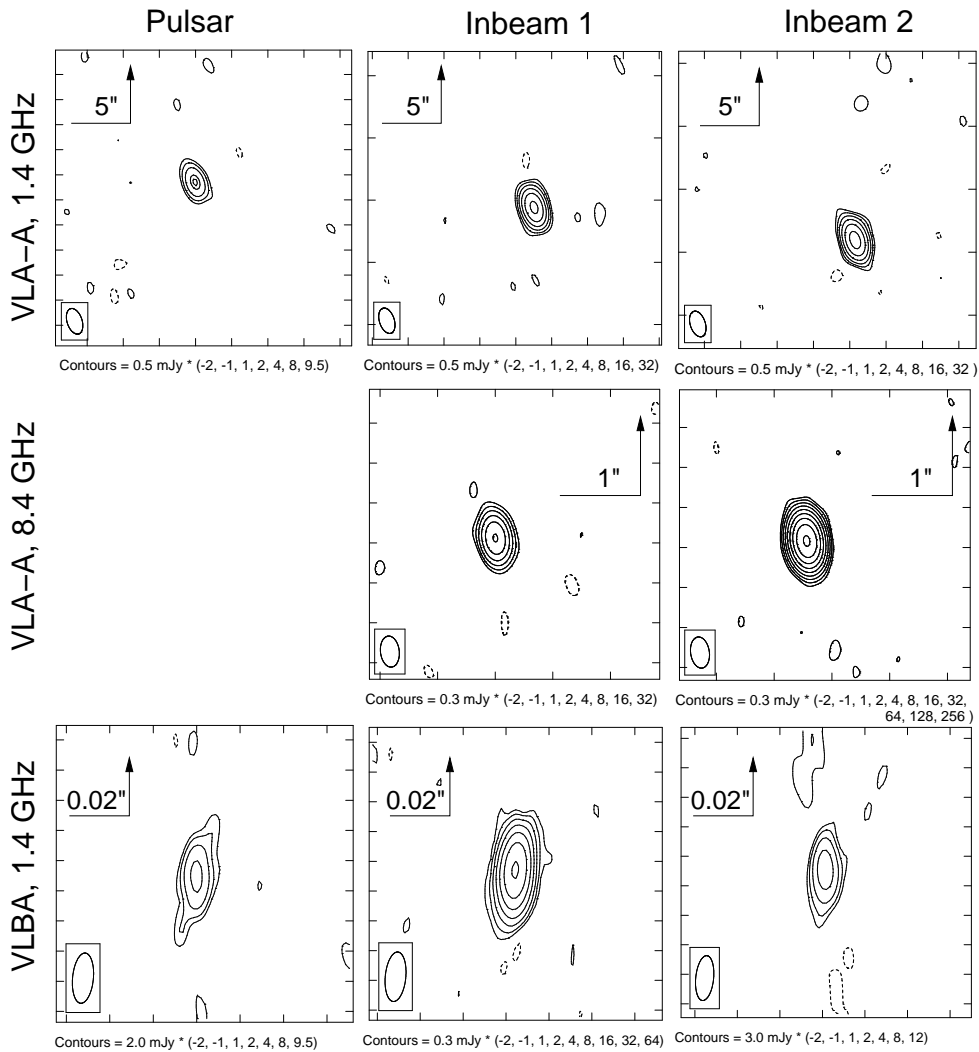}
\caption{
An illustration of the selection and refinement process for
  in-beam calibrator candidates for PSR B2045$-$16. Note the change
  in field of view for each row.
Top row: Images at 1.4~GHz, VLA A-configuration, showing the pulsar
 and two compact in-beam sources. The beam size is $\sim$1\farcs4, and
 the image RMS noise level is $\sim$0.5~mJy/beam. Besides the pulsar,
 14 sources were detected in the field, and 6 were found to be
 relatively compact.
Middle row: Follow-up observations at 8.4~GHz, VLA A-configuration.
 The beam size is $\sim$0\farcs2, and the image RMS noise level is
 $\sim$0.3~mJy/beam.  The pulsar was not observed at this high
 frequency, but all 6 targets were; 2 of the 6 (shown here) are
 observed to be compact.
Bottom row: First epoch VLBA observations at 1.4 GHz. The beam size is
 $14\times5$~mas, and the ungated image RMS noise level is
 $\sim$0.3~mJy/beam.  The pulsar and both in-beam calibrator
 candidates are detected.  
Note that for plotting purposes only, the flux densities of the
 in-beam candidates at the VLA, 1.4~GHz are not corrected for primary
 beam attenuation, and so apparently increase in the pointed VLBA
 1.4~GHz observations, even though in practice some of the source flux
 density is probably resolved out.
\label{Fig:inb}
}
\end{figure*}

\subsection{Final Target Selection}

From the original sample of 63 pulsars, 34 were selected for
observations with the VLBA, most of them with good candidate in-beam
calibrators.  A few pulsars were judged to be strong enough and to
have a cataloged VLBA calibrator that was close enough that phase
referencing, in combination with wideband ionospheric calibration
\citep{bbgt02}, would provide accurate astrometry. These pulsars will
be discussed in future work.

In the first epoch of VLBA observations, most of the selected pulsars
were found to have at least one suitable in-beam calibrator, as
illustrated in \Fref{inb}. However, the in-beam candidates for 7
pulsars were either not detected or were too heavily resolved to be
useful.  In addition, 3 pulsars were not detected in first epoch VLBA
observations due to scattering or poor calibration at lower
declinations; these were discarded from our observing list, and
substituted with other targets.

The process outlined here was optimized to identify pulsars where a
parallax measurement seemed plausible based on our previous
experience.  No attempt was made to preserve sample completeness or to
image each possible in-beam calibrator candidate exhaustively, and as
such, it is difficult to infer detailed statistical properties for the
ensemble of faint, compact background sources.  Of the 97 sources that
were found to be compact in both 1.4~GHz and 8.4~GHz VLA
A-configuration observations, the vast majority (75 sources) had a
negative spectral index ($S_\nu \propto \nu^{\alpha}$; $\alpha < 0$),
as expected for non-thermal sources.  (The spectral index was
determined after correcting measured flux densities at 1.4~GHz for
primary beam attenuation.)
Specifically, $\alpha < -0.5$ for 43 sources, and $\alpha < -1.0$ for
16 sources.  We emphasize that these results do not necessarily
contradict the notion that compact sources tend to have flat spectral
indices \citep[e.g.,][]{pmj83}.  Our sample is faint, and likely to
include a mixture of different kinds of objects, including some
core-jet sources.  Additionally, our two-point spectral index might
straddle a peak in the spectrum, producing an apparent distribution of
spectral indices that is hard to interpret.  \citet{wjs+05} and
\citet{fkc+06} find broadly similar spectral index distributions for
their source samples.

\section{VLBA Observations}\label{Sec:vlbaobs}

While pulsar astrometric observations have previously been conducted
on a number of VLBI networks, the VLBA has several features that are
particularly well suited to astrometry.  The identical dishes of the
VLBA simplify the calibration and ensure control of systematic errors,
and the identical primary beams (as well as the match with the primary
beam of the VLA dishes) is particularly advantageous for in-beam
calibration.  In a heterogeneous array, in-beam calibrators would have
to be restricted to the field of view of the largest telescope in the
network, which can severely restrict the chances of finding a suitable
calibrator.  Secondly, full-time operation and flexible dynamic
scheduling allow several short observations of each target through the
year, timed to track parallax extrema and to avoid observations
through the solar corona.  Thirdly, the VLBA was designed for phase
referenced observations with rapid slews and short dwell times, and
provides repeatable phase stability, an essential component for
astrometry.  Finally, pulsar astrometry in particular can take
advantage of pulsar gating, where the VLBA correlator effectively
accumulates signal only when the radio pulse is predicted to arrive
and discards the noise at other times.

\subsection{Observing Strategy and Scheduling}\label{Sec:vlbsched}

Our choice of observing frequency balances competing priorities.
Radio pulsars typically have steep spectra ($S_\nu \propto
\nu^\alpha$; $\alpha \sim -1.6$), and thus are brighter at lower
frequencies.  Additionally, the larger field of view at lower
frequencies enhances the likelihood of finding in-beam calibrators,
and tropospheric phase fluctuations are reduced at lower frequencies.
On the other hand, the synthesized beam is broader, resulting in lower
astrometric precision when sensitivity is the limiting factor (the
case for many of the targets and calibrators here), and 
the phase distortions due to the
differential ionosphere between the target and the calibrator increase
rapidly at lower frequencies.  An observing frequency around 1.5~GHz
was chosen as a compromise between these effects, as demonstrated in
our various pilot projects \citep{bbb+00,ccl+01}; the residual
differential ionospheric effects remain the largest source of
astrometric uncertainty in our program.

In an effort to minimize systematic errors between epochs, all of the
observations were set up in as identical a manner as possible.  Dual
circular polarization was recorded simultaneously in four separate
8-MHz spectral windows.  These bands were Nyquist sampled with 4-level
quantization.  We chose widely spread spectral windows to enable
calibration of the ionosphere \citep{bbb+00} for cases where in-beam
calibration proved difficult and the pulsar was deemed bright enough
to allow this method to succeed.  The in-beam calibration technique
described in the present work does not make use of the wide-spread
bands.  The initial set of spectral windows was
centered on 1414, 1489, 1594, 1684~MHz, choices that were informed by
presence of known satellite-generated interference.  After routinely
encountering strong radio frequency interference (RFI), the second
spectral window was changed to a central frequency of 1508~MHz.  These
spectral windows proved to be relatively RFI-free for most of the
observations, although strong narrow-band RFI was occasionally
observed around 1684~MHz, mainly at the Mauna Kea site.  The VLBA
pulse calibration system was turned off during these observations in
order to reduce the total system temperature by about 3\%.

At each epoch, an observation cycled between the field containing the
pulsar and its in-beam calibrator(s) and a VLBI reference calibrator.
The VLBI reference calibrators, which are listed in \Tref{cals}, were
drawn from the standard VLBI catalogs \citep{vcs1,vcs2,vcs3} and are
the ultimate means by which the observations are tied to the {ICRF}.
The calibrator was chosen primarily for proximity to the target
pulsar, though strength and compactness requirements drove some
choices.  The pointing direction for a target field was set to be
between the pulsar and its in-beam calibrator, with the restriction
that the offset from the pulsar be no more than 6\arcmin; for some
pulsars, two in-beam calibrators could be identified and the pointing
was set to lie between all three sources in the field.  Each
observation spanned a duration of 2~hours approximately centered on
the transit time of the source at the center of the VLBA.  We used
interleaved scans of
90~seconds on the reference calibrator and 120~seconds on the target
field, resulting in a total integration time of about 1~hour on the
target field after accounting for slewing overheads.

\begin{deluxetable*}{lllll}
\tablecolumns{6}
\tablewidth{0pc}
\tablecaption{List of Calibrators\label{Tab:cals}}
\tablehead{
  \colhead{Pulsar} & \colhead{Calibrator} & \colhead{R.A.}
& \colhead{Dec.} & \colhead{Reference}  \\
  \colhead{} & \colhead{} & \colhead{(J2000)}
& \colhead{(J2000)} & \colhead{}
}
\startdata
B0031$-$07 &	J0024$-$0412 &		00:24:45.983228 &	$-$04:12:01.54874 &	VCS	\\
	 &	In-beam  &		00:33:27.77786  &	$-$07:12:38.9802  &		\\
B0136$+$57 &	J0147$+$5840 &		01:47:46.541149 &	$+$58:40:44.97198 &	VCS	\\
	 &	In-beam &		01:37:50.47629  &	$+$58:14:11.2905  &		\\
B0450$-$18 &	J0450$-$1837 &		04:50:35.909627 &	$-$18:37:00.40798 &	VCS	\\
	 &	In-beam &		04:52:11.97908  &	$-$18:16:41.9582  &		\\
B0450$+$55 &	J0458$+$5508 &	        04:58:54.839957 &	$+$55:08:42.05815 &	VCS	\\
J0538$+$2817 &	J0547$+$2721 &		05:47:34.148923 &	$+$27:21:56.84262 &	ICRF	\\
	 &	In-beam &		05:38:16.8559   &	$+$28:24:29.4778  &		\\
B0818$-$13 &	J0820$-$1258 &		08:20:57.447616 &	$-$12:58:59.16914 &	ICRF	\\
	 &	In-beam &		08:20:11.43071  &	$-$13:49:12.8558  &		\\
B1508$+$55 &	J1510$+$5702 &		15:10:02.922368 &	$+$57:02:43.37587 &	ICRF	\\
	 &	In-beam &		15:11:48.06076  &	$+$55:41:56.2876  &		\\
B1541$+$09 &	J1555$+$1111 &		15:55:43.044016 &	$+$11:11:24.36571 &	VCS	\\
	 &	In-beam &		15:43:03.70214  &	$+$09:19:09.7845  &		\\
J1713$+$0747 &	J1719$+$0817 &		17:19:52.206216 &	$+$08:17:03.55378 &	VCS	\\
	 &	In-beam &		17:13:27.09521  &	$+$07:30:18.9931  &		\\
B1933$+$16 &	J1924$+$1540 &		19:24:39.455870 &	$+$15:40:43.94172 &	VCS	\\
	 &	In-beam &		19:35:55.52636  &	$+$16:17:00.6849  &		\\
B2045$-$16 &	J2047$-$1639 &		20:47:19.667027 &	$-$16:39:05.84254 &	VCS	\\
	 &	In-beam &		20:48:49.57191  &	$-$16:00:13.9195  &		\\
B2053$+$36 &	J2052$+$3635 &		20:52:52.054980 &	$+$36:35:35.30036 &	VCS	\\
	 &	In-beam &		20:55:12.22739  &	$+$36:34:22.6929  &		\\
B2154$+$40 &	J2202$+$4216 &		22:02:43.291372 &	$+$42:16:39.97992 &	ICRF	\\
	 &	In-beam &		21:57:54.02421  &	$+$40:04:54.6025  &		\\
B2310$+$42 &	J2322$+$4445 &		23:22:20.358088 &	$+$44:45:42.35349 &	VCS 	\\
	 &	In-beam &		23:12:13.49560  &	$+$42:39:32.3585  &		\\
\enddata
\tablecomments{Primary and in-beam calibrators observed for astrometry
  of pulsars.  The primary calibrators are either part of the ICRF
  \citep{mae+98,fma+04} or tied to the ICRF by the VLBA Calibrator
  Survey programs \citep[VCS;][]{vcs1,vcs2,vcs3}.  When in-beam
  sources are listed, the pulsar positions are tied to the positions
  of the primary calibrator through the in-beam sources.  All in-beam
  calibrators were discovered as part of this project, as described in
  \Sref{inb}. ICRF sources used here have positional uncertainties of
  $\sim 200~\mu$as.  The sources from the VCS catalog have
  positions that are certain to 5~mas or better. }
\end{deluxetable*}

We aimed to observe each pulsar for a total of 8~epochs over the
course of 2~years, with epochs spaced approximately 3~months apart.
Observations for a given pulsar were conducted at a similar LST at
each epoch in order to provide nearly identical spatial frequency
sampling (``$U-V$ coverage'').  The observations were
typically scheduled within 2 weeks of parallax extrema.
Between~2002 June~27 and~2005 March~25, our observing programs (BC120,
BC123) accumulated a total of 278 observation blocks of two hours each.

A few observations failed and had to be discarded due to the proximity
of the \objectname[]{Sun} to the target pulsar in the sky, a
circumstance whose severity was not fully appreciated by us when the
program began.  Turbulence in the extended solar corona and the solar
wind can cause phase fluctuations \citep{skk+02} which degrade
astrometry.  If the target field was within $\sim$10\arcdeg\ of the
Sun, the images of the calibrators were severely distorted, but the
solar corona and wind can affect 1.5~GHz observations of targets as
far away as 45\arcdeg\ from the Sun.  In order to avoid severe phase 
errors, future astrometric observations should not be scheduled
within about~30\arcdeg\ of the \objectname[]{Sun} at 1.5~GHz.

\subsection{Pulsar Timing and Gating}\label{Sec:gate}

\emph{A priori} information about the rotational phase of a pulsar can
be exploited to correlate the observations only when the pulsar is
``on'' and the pulse is visible, while discarding data from the
remainder of each pulse period.  The result is a S/N improvement
scaling as $\approx \sqrt{P/w}$ for a pulsar with pulse period~$P$ and
pulse width~$w$, with typical improvement factor $\sim$3.  The
ephemerides used for gating the pulse signal were obtained from timing
observations with the 76-m Lovell telescope of the Jodrell Bank
Observatory.  Typically, observing spans of a few months and
overlapping the VLBA observation epoch were used to refine the pulsar
rotational parameters in order to account for the effects of long-term
timing noise.

The pulsar timing observations were made primarily at frequencies
around 1.4~GHz, employing a dual-channel cryogenic receiver system
sensitive to two orthogonal circular polarizations. The signals from
each polarization were mixed to an intermediate frequency, fed through
a multichannel filterbank, and digitized.  The data were dedispersed
in hardware and folded on-line according to the dispersion measure and
topocentric period of the pulsar.  The folded pulse profiles were
stored for subsequent analysis.

Complete details regarding the procedure used to determine the pulsar
rotational phases can be found in \citet{hlk+04}; here we summarize
only the salient aspects.  Each timing observation consisted of a
series of 1--3~minute integrations, with a total duration of up to
$\sim$45~minutes.  The total observing time per pulsar was determined
by the amount of time required to obtain sufficient sensitivity, given
the pulsar flux density.  In post-processing, any of the individual
integrations that appeared to be dominated by RFI were removed, the
polarizations combined, and the data were averaged to produce a single
total-intensity profile for the observation.  Pulse times of arrival
(TOAs) were subsequently determined by finding the peak in a time
domain convolution of the averaged profile with a pulse profile
template corresponding to the observing frequency.  These TOAs were
corrected to the solar system barycenter using the Jet Propulsion
Laboratory DE200 solar system ephemeris \citep{s82}, analyzed using
the standard TEMPO software package\footnote{{\tt
    http://www.atnf.csiro.au/research/pulsar/tempo/}}, and transmitted
to the VLBA correlator for use in the correlation.  We note that our
refined astrometry does not affect the pulse phase prediction from the
timing solutions at a relevant level, and while gating improves the
S/N ratio of the pulsar in VLBA observations, it does not otherwise
influence the astrometry.

\subsection{Correlation}\label{Sec:correlate}

In order to mitigate time-average and bandwidth smearing (as for the
{VLA}, \Sref{survey}), the correlation was performed with~32
spectral channels of~250~kHz width and then averaged for~2~s.  Both
time-average and bandwidth smearing should be negligible over a
30\arcmin\ (FWHM) field of view; however, the ionosphere can induce
unpredictable delays and rates that could otherwise reduce the 
correlation coefficients through decorrelation for greater degrees 
of averaging. Only the parallel senses of polarization were
correlated.

Each of the in-beam calibrators and pulsars were correlated in
separate passes, with the phase reference center set in turn to the
in-beam calibrator or pulsar positions.  The pulsar correlation pass
was gated as described above.  Correlator parameters, such as clock
offsets, station locations, and Earth orientation parameters, were
kept identical to enable cross calibration between the correlator
passes.

\section{VLBA Data Reduction}\label{Sec:reduce}

The multi-epoch observations of each pulsar typically produced 18~TB
of raw voltage-sampled data that generated 5~GB 
of correlated data, from which 5 astrometric parameters ($\alpha_0$,
$\delta_0$, $\mu_{\alpha}$, $\mu_{\delta}$, $\pi$)\footnote{In this 
work, both $\mu_{\alpha}$ and $\mu_{\delta}$ always refer to angular,
rather than coordinate, rates.  Specifically, $\mu_{\alpha} =
\frac{d\alpha}{dt}\cos \delta$.} are finally
derived; the factor of~$\sim 10^{12}$ truly warrants the use of the
term ``data reduction'' for VLBI pulsar astrometry.  Given the data
volume, a semi-automated pipeline was built within the NRAO
Astronomical Image Processing System (AIPS\footnote{\tt
http://www.aips.nrao.edu}) for calibration and imaging.  The pipeline
allowed for human interaction during data editing and specific imaging
steps while minimizing errors during repetitive book-keeping and
calibration steps.  We summarize the various calibration steps below,
with emphasis on the specific corrections introduced.

\subsection{{\em A priori} Calibration}\label{Sec:apriori}

Corrections for various instrumental and system parameters are
determined from geometric factors and ancillary data. Since they do
not require observations of an astronomical source, we distinguish
them as {\em a priori} calibration, following usual practice.  These
corrections are listed below, in the order in which they were applied.

{\em Earth Orientation Parameters:} In order to combine the signals
from the various antennas, the correlator must have an accurate
geometrical model of the array \citep{r99}. The model includes the
orientation of the Earth in space, as described by a set of Earth
orientation parameters (EOPs). The most accurate estimates of these
EOPs are typically available only after a time lag of several weeks,
well after data correlation.  Moreover, during the course of this
project, it was discovered that the VLBA correlator had inadvertently
been provided with predicted rather than measured EOPs 
(Walker et al.\ 2005, VLBA Test Memo\footnote{See {\tt
http://www.vlba.nrao.edu/memos/test/}} 69).  The resultant
differential astrometric errors are potentially as large as $\sim
2$~mas, dwarfing the astrometric error budget.  The correction for this
was a delay offset computed between the delay model using the
(incorrect) correlation EOPs and the actual EOPs, and this correction
was applied to each data set.

{\em Antenna Positions:} The locations of the VLBA antennas have an
impact on absolute astrometry at the level of $\sim 50$~\muas\ per mm
of position error on the longest baselines; the astrometric consequences
are worse on shorter ones.  The errors in differential astrometry are smaller
than this by roughly the target-calibrator separation (in radians) or
$\sim 1$~\muas\ per mm of position error per degree of separation on
the sky.  On 2004~May~12, a new model for antenna positions and
velocities (GSFC solution 2004b) was adopted at the VLBA correlator,
introducing a typical discontinuity of $\sim 20$~mm in antenna
positions.  Each data set correlated before the change was corrected
by shifting antenna positions to be consistent with the 2004b
solution.

{\em Quantization Corrections:} Because the VLBA uses two- or
four-level quantization when sampling, the visibilities produced by the
VLBA correlator differ systematically from those that would be
produced by an ideal analog system.  Quantization statistics were
calculated for the auto-correlations and used to correct the
amplitudes of the cross-correlations.

{\em Gain and System Temperature:} The measured system temperatures
and antenna gains (typically $\sim$0.1~K~Jy$^{-1}$ for a VLBA
antenna at 1.5~GHz) were used to scale the correlation
amplitudes, in the standard manner.

{\em Parallactic Angle Correction:} The VLBA antennas have
altitude-azimuth mounts, so the antenna feeds rotate as the antennas
track sources across the sky.  Corrections were applied for the
geometric phase shift introduced between the left and right circularly
polarized signals due to feed rotation.

{\em Coarse Ionospheric Correction:} Precision astrometry requires
correction for the differential phase and delay effects introduced by
signal propagation through the Earth's ionosphere.  The effects due to
different ionospheric conditions above each antenna fall into two 
categories: image distortion and image shift.  The former causes an 
overall decrease in the signal-to-noise of the astrometry by broadening
the image.  The latter introduces an unknown, unwanted position offset, 
directly impacting the astrometric results.

The ionospheric total electron content (TEC) can be estimated by
measuring the propagation delay between Global Positioning System
(GPS) signals at two different frequencies along the lines of sight to
the constellation of GPS satellites.  Global maps of the ionospheric
TEC are routinely produced based on such measurements from around the
world, and archived by the NASA Crustal Dynamics Data Information
System (CDDIS\footnote{{\tt http://cddisa.gsfc.nasa.gov/}}).  These
maps are relatively coarse, typically quantifying the ionospheric TEC
on a grid with pixels spaced 5\arcdeg\ in longitude and 2\fdg5 in
latitude, and updated every 2~hr.  While these coarse models are far
from sufficient for precision astrometry, Walker \& Chatterjee (1999,
VLBA Scientific Memo\footnote{See {\tt
    http://www.vlba.nrao.edu/memos/sci/}} 23) use them to find a
reduction in phase delay scatter by a factor of 2 to 5 at 2.2~GHz, and
using them to correct for ionospheric dispersion delay did improve
phase referencing.

\subsection{Radio Frequency Interference Identification and Excision}
\label{Sec:edit}

Even though the spectral windows were chosen to minimize the overlap
with known sources of interference, some residual RFI persisted in the
data.  While the usual interferometric practice is to edit data on a
per-baseline basis, we concluded that RFI could be identified on an
\emph{antenna} basis, i.e., identified with being present only on
baselines common to a specific antenna.  Therefore, we could identify
and flag RFI based on the autocorrelation data at each antenna, and
baseline-based RFI excision was only rarely needed.

Data inspection and editing required extensive human intervention.
However, since the same data were correlated multiple times, we could
identify and flag RFI on the output of the first correlation pass and
apply the flags to visibilities generated on all further passes.

\subsection{Astronomical Calibration}\label{Sec:calibrate}

Phase, delay, and bandpass calibrations for VLBI are typically derived
from observations of calibrator sources.  For a given pulsar the same
bright VLBA calibrator, as listed in \Tref{cals}, was used for all
three purposes, and we followed an iterative process of calibration
refinement.  The calibration steps listed below were all performed
using AIPS.

{\em Fringe fitting:} We solved for calibrator visibility phase delays
and time derivatives (rates) by fitting for a phase slope across each
frequency band, assuming a point source model. The delays were
smoothed by averaging over 15 minutes and interpolated for
observations of the targets.

{\em Bandpass calibration:} After fringe-fitting, calibrator data from
all epochs on a particular target were combined, and the inner 26 of
the 32 channels in each frequency band were then imaged. Several
iterations of imaging and self-calibration were performed to create a
final calibrator image, averaged over all observation epochs.  The
resulting `global' calibrator model was used for bandpass calibration
at each epoch, determining a phase and amplitude gain factor for each
of the 32 channels of each frequency band, for each antenna.  These
corrections help to minimize baseline-dependent errors in further
calibration steps.

{\em Frequency division:} The calibrator structure did not change
significantly over the duration of the project, enabling us to build
time-averaged models.  However, in some cases the calibrator images
showed significant frequency dependent structure.  As such, after the
initial fringe-fitting and bandpass calibration, we used separate
calibrator models for each frequency band.

{\em Phase and amplitude calibration:} For each frequency band,
calibrator data from all epochs were combined, and an imaging and
self-calibration loop was used to derive an epoch-averaged calibrator
model.  The global calibrator model produced for bandpass calibration
was used as the initial model in each case.  Phase and amplitude
calibrations were then derived independently for each epoch and
frequency band, using calibrator models that were averaged over all
epochs but not over frequencies.  The phase and amplitude calibration
was then applied to the target source (i.e., either the pulsar or the
in-beam calibrator) and written to a single-source, calibrated UV
database.  In order to ensure that the data for the pulsar and in-beam
calibrator(s) were identically calibrated, we derived the
fringe-fitting, bandpass calibration, and phase/amplitude calibration
for one correlation pass and copied these tables to all subsequent
correlation passes.

At the end of this calibration sequence there were typically
32~calibrated single-source UV databases per pulsar and per in-beam
calibrator, one for each of 4~frequency bands at 8~epochs.

\subsection{In-beam calibration}

The angular separation between an in-beam calibrator and the target
pulsar (typically $\sim$15--20\arcmin) is much smaller than their
angular separation from the phase calibrator (typically
$\sim$2--3\arcdeg). As such, most of the residual calibration errors
will be common to the pulsar and the in-beam source, and we can
exploit in-beam calibrators to incrementally improve the calibration
of the pulsar.

As for the phase calibrators, we created independent models for
in-beam calibrators at each frequency band by combining and imaging
the calibrated visibility data from all epochs at that band. A few
passes of self-calibration (usually phase-only, but including
amplitude self-calibration for in-beam sources brighter than $\sim
50$~mJy) were used to improve the models.  Since in-beam calibrators
are usually fairly weak ($\sim 10$~mJy) and show some source
structure, we required longer time averages and lower S/N cutoffs for
the calibration solutions, up to a limit of 5~minutes of averaging and
S/N $\gtrsim 1.8$. This lower S/N threshold was determimed to maximize
the S/N in the phase-referenced image of the pulsar after the phase
transfer from the in-beam calibration solutions; calibrators with
different characteristics may be optimized with slightly different
cutoff values.  In some cases, the in-beam source had enough
resolved structure that the longest baselines of the VLBA (to Saint
Croix and Mauna Kea) had to be discarded.  Even in such cases, using
in-beam calibration provided superior astrometric results compared to
retaining the longest baselines while only referencing observations to
the more distant phase calibrator.

With an epoch-averaged model for the in-beam calibrator at each
frequency band, we used a final pass of phase-only self-calibration to
obtain phase corrections.  These corrections were applied to the
calibrated pulsar data and used for subsequent imaging.

\subsection{Imaging and position fitting}

Since the target pulsars are intrinsically compact, the final imaging
step is quite simple.  The VLBA beam is North-South elongated, with
typical dimensions of $5 \times 14$~mas for our chosen frequencies,
observation durations, and selection of source declinations.  
We used pixel sizes of $1 \times 1$~mas and
approximately natural weighting (an AIPS `robustness' parameter of
$2$, where $5$ corresponds to pure natural weighting and $-5$
corresponds to uniform visibility weights; see \citealt{bss99}).  Many
of the target pulsars have observed diffractive scintillation
timescales at 1.5~GHz that are comparable to the 2~hr observation
span, and so their sidelobes are often not well subtracted.  For this
reason support constraints (i.e. `boxes') were used to restrict the
region to be cleaned, and we stopped the cleaning process
interactively once the residuals within the support box were at the
level of the image noise far from the pulsar.  Finally, we determined
astrometric positions for our targets at each epoch and frequency band
by fitting elliptical Gaussians in the image domain.

\section{Analysis and Results}\label{Sec:analysis}

At the end of the calibration and imaging process, we are left with a
set of positions $\{\alpha_{t,\nu}, \delta_{t,\nu};\; t=1 \ldots 8,\,
\nu=1 \ldots 4\}$ at each epoch $t$ and frequency band $\nu$.  We fit
these positions with a five-parameter astrometric model encompassing
position $(\alpha_0,\delta_0)$, proper motion
$(\mu_\alpha,\mu_\delta$), and parallax $(\pi)$ using a standard
weighted linear least squares technique.  The key problem is the
determination of the weights, and hence the fit uncertainties, in the
presence of unknown systematic errors at each epoch.

At each epoch, the target position errors have a (random) measurement
error component, corresponding to the beam FWHM/$(2 \times S/N)$,
where $S/N$ is the signal-to-noise ratio of the final pulsar image.
However, using only these random errors produces fits with reduced
$\chi^2$ values $>1$, implying the presence of unmodeled systematic
errors.  In VLBI astrometry at 1.5~GHz, the biggest systematic error
component is usually due to the residual unmodeled ionosphere between
the in-beam calibrator and the target pulsar, with other possible
contributions from errors in the Earth orientation parameters
(c.f.~\Sref{apriori}), calibrator position errors, or structural
evolution of the calibrators.  The calibrator position errors could in
principle impart a significant ($\sim 50~\mu$as) time-dependent error
on the phase-referenced position measurements.  (We note that
uniformly observing a given target over the same hour angle range
causes calibrator source position errors to produce a consistent
astrometric offset at each epoch, affecting only the absolute position
determination.)  A common approach to treating unmodeled
errors is to inflate all the random errors by a trial factor until a
reduced $\chi^2$ value $\approx 1$ is obtained.  However, the residual
ionospheric errors, for example, depend on the proximity of the Sun
(c.f.~\Sref{vlbsched}), and since the separation varies with season,
scaling the random errors at each epoch by the same factor would not
be optimal.  In cases where the angular separation between the sun and
the target is more than 30\arcdeg, the uncalibrated ionosphere typically
yields astrometric errors of $\sim 200~\mu$as for a 1\arcdeg\  
target-calibrator separation at an observing frequency of 1.5~GHz.

In our past programs \citep[e.g.,][]{ccl+01,ccv+04} we have
investigated several possible measures of the epoch-dependent
systematic error, including the angular size of the pulsar image and
its deviations from the beam shape, or the scatter between the
positions determined at each frequency band at a given epoch.  While
these measures are useful, they are not perfect, and regression
analysis between the estimated systematic error and parameters such as
the Sun angle or the elevation of the source above the horizon do not
reveal a simple correlation.

Instead we choose to estimate the fit uncertainties from the position
measurements themselves using the bootstrap technique
\citep[e.g.,][]{et91}.  We treat the $N_d$ data points
$\{\alpha_{t,\nu}, \delta_{t,\nu}\}$ at each epoch and frequency band
as independent estimates of the (time dependent) position, and select
$N_d$ samples from the set {\em with replacement}, discarding
degenerate samples with fewer than three independent positions. We
then fit the sample for astrometric parameters using standard weighted
least squares, using only the random measurement uncertainties.  The
process of sampling and fitting is repeated a large number of times.
Each resulting fit generally has a reduced $\chi^2 >1$, but only the
values of the fit parameters are of interest.  Histograms of the
resulting fit parameters are shown for a typical case with 50,000
sampling iterations (PSR~B2310+42, \Fref{boot1}).  It is apparent that
the parameters follow a smooth distribution with a well-defined most
probable value (the distribution mode). The most probable value for
each astrometric parameter and their most compact 68\% confidence
intervals as estimated from the bootstrap method are listed in
\Tref{results} for the fourteen pulsars in our final sample.  The most
probable values and confidence intervals for the distances and
transverse velocities of these pulsars were also derived directly via
bootstrap, and are listed in \Tref{derivpars}.

\begin{figure}[t]
\epsscale{0.9}
\plotone{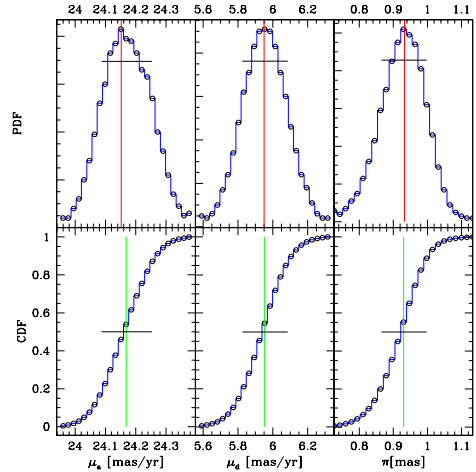}
\caption{Uncertainties in astrometric parameters as estimated by the
  bootstrap method for PSR~B2310+42. The top row shows histograms of
  the estimates for the fit parameters from 50,000 bootstrap
  iterations, equivalent to an un-normalized probability distribution
  function (PDF), and the vertical line indicates the most common value
  (the mode). The bottom row shows the cumulative distribution
  function (CDF) derived from the histogram above, and the vertical line
  indicates the median parameter value. In each panel, the horizontal
  bar indicates the most compact 68.26\% confidence interval (i.e., the
  smallest interval that includes 34,130 estimates for each given
  parameter). In this and every other case, the median and mode are
  consistent to well within the 68\% confidence interval.
  The proper motion in right ascension, $\mu_{\alpha}$, is the angular rate,
  $\frac{d\alpha}{dt}\cos \delta$.
\label{Fig:boot1}
}
\end{figure}

\begin{figure}[t]
\epsscale{0.9}
\plotone{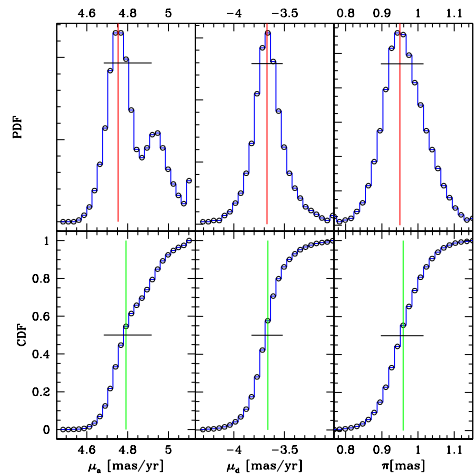}
\caption{Uncertainties in astrometric parameters as estimated by the
  bootstrap method for PSR~J1713+0747. As in \Fref{boot1}, the
  top row shows the histogram of the estimates (equivalent to an
  un-normalized PDF) and the bottom row shows the cumulative
  distribution function (CDF) for each astrometric parameter, while
  vertical lines indicate the mode and median estimates and horizontal
  bars show the most compact 68\% confidence interval for each
  parameter. The fit for $\pmra$ is the only case among all pulsars
  discussed in the present work where a bimodal distribution was
  obtained for an astrometric parameter, and thus represents our
  worst-case scenario. However, our astrometric results for this
  pulsar can be compared to independent estimates from pulse timing:
  see \Sref{frametie} and \Tref{J1713} for a discussion. 
\label{Fig:boot2}
}
\end{figure}

In all cases, the values of the parameters estimated by the bootstrap
method were consistent with the parameters obtained by simple least
squares fitting (using all epochs without replacement) to well within
the bootstrap 68\% confidence interval.  
However, the uncertainties in the estimates of
the astrometric parameters differed between the two methods.  The
power of the technique is illustrated by our worst-case scenario
(PSR~J1713+0747, \Fref{boot2}), which was the only case where a
bimodal distribution was obtained for any of the parameters. The
bimodality in $\mu_\alpha$ reflects the presence or absence of one
well-measured position, and causes the uncertainties to deviate
significantly from symmetry.  However, there is no ambiguity in the
most probable parameter values or in the bootstrap confidence
intervals.  Coincidentally, PSR~J1713+0747 has well determined
astrometry from pulse timing as well, allowing a stringent check on
our results, as we discuss further below (\Sref{frametie}).

\begin{deluxetable*}{lllrrr} 
\tablecolumns{6}
\tablewidth{0pc}
\tablecaption{Astrometry Results\label{Tab:results}}
\tablehead{
  \colhead{Pulsar} & \colhead{RA} & \colhead{Dec} 
& \colhead{$\pmra$} & \colhead{$\pmdec$} & \colhead{$\pi$} \\
  \colhead{} & \colhead{(J2000)} & \colhead{(J2000)} 
& \colhead{(\mpy)} & \colhead{(\mpy)} & \colhead{(mas)} 
}
\startdata 
     B0031$-$07 &  
$  00:34:08.8703(1) $ & $  -07:21:53.409(2) $ & $  10.37^{+0.08}_{-0.08}$  &  $ -11.13^{+0.11}_{-0.16}$  &  $   0.93^{+0.08}_{-0.07}$\\
   B0136+57  &  
$  01:39:19.7401(12) $ & $  +58:14:31.819(17) $ & $ -19.11^{+0.06}_{-0.07}$  &  $ -16.60^{+0.06}_{-0.07}$  &  $   0.37^{+0.04}_{-0.04}$\\
  B0450$-$18  &  
$  04:52:34.1057(1) $ & $  -17:59:23.371(2) $ & $   8.89^{+1.60}_{-2.24}$  &  $  10.65^{+1.87}_{-1.29}$  &  $   0.65^{+1.40}_{-0.59}$\\
     B0450+55  &  
$  04:54:07.7506(1) $ & $  +55:43:41.437(2) $ & $  53.34^{+0.06}_{-0.05}$  &  $ -17.56^{+0.12}_{-0.14}$  &  $   0.84^{+0.04}_{-0.05}$\\
     J0538+2817  & 
$  05:38:25.0572(1) $ & $  +28:17:09.161(2) $ & $ -23.57^{+0.10}_{-0.10}$  &  $  52.87^{+0.09}_{-0.10}$  &  $   0.72^{+0.12}_{-0.09}$\\
     B0818$-$13  &  
$  08:20:26.3817(1) $ & $  -13:50:55.859(2) $ & $  21.64^{+0.09}_{-0.09}$  &  $ -39.44^{+0.04}_{-0.05}$  &  $   0.51^{+0.03}_{-0.04}$\\
    B1508+55  &  
$  15:09:25.6298(1) $ & $  +55:31:32.394(2) $ & $ -73.64^{+0.05}_{-0.04}$  &  $ -62.65^{+0.09}_{-0.08}$  &  $   0.47^{+0.03}_{-0.03}$\\
     B1541+09  &  
$  15:43:38.8250(1) $ & $  +09:29:16.339(2) $ & $  -7.61^{+0.06}_{-0.05}$  &  $  -2.87^{+0.06}_{-0.07}$  &  $   0.13^{+0.02}_{-0.02}$\\
     J1713+0747  &  
$  17:13:49.5306(1) $ & $  +07:47:37.519(2) $ & $   4.75^{+0.17}_{-0.07}$  &  $  -3.67^{+0.16}_{-0.15}$  &  $   0.95^{+0.06}_{-0.05}$\\
     B1933+16  &  
$  19:35:47.8259(1) $ & $  +16:16:39.986(2) $ & $   1.13^{+0.12}_{-0.13}$  &  $ -16.09^{+0.15}_{-0.13}$  &  $   0.22^{+0.08}_{-0.12}$\\
     B2045$-$16  &  
$  20:48:35.6013(2) $ & $  -16:16:44.534(3) $ & $ 113.16^{+0.02}_{-0.02}$  &  $  -4.60^{+0.28}_{-0.23}$  &  $   1.05^{+0.03}_{-0.02}$\\
   B2053+36  &  
$  20:55:31.3521(1) $ & $  +36:30:21.469(2) $ & $   1.04^{+0.04}_{-0.04}$  &  $  -2.46^{+0.13}_{-0.13}$  &  $   0.17^{+0.03}_{-0.03}$\\
  B2154+40  &   
$  21:57:01.8495(1) $ & $  +40:17:45.986(2) $ & $  16.13^{+0.09}_{-0.10}$  &  $   4.12^{+0.12}_{-0.12}$  &  $   0.28^{+0.06}_{-0.06}$\\
     B2310+42  & 
$  23:13:08.6209(1) $ & $  +42:53:13.043(2) $ & $  24.15^{+0.10}_{-0.07}$  &  $   5.95^{+0.13}_{-0.12}$  &  $   0.93^{+0.06}_{-0.07}$\\
\enddata
\tablecomments{Pulsar coordinates have been adjusted for improvements
  in primary calibrator positions.  Uncertainties in these coordinates
  (in parentheses) reflect uncertainties in the positions of primary
  calibrator; a minimum uncertainty of 0.0001 second of time in right
  ascension and 2~mas in declination are assigned as calibrator source
  structure evolution with frequency could cause systematic offsets
  from the reported 8.4~GHz positions at about this level.  In all
  cases this calibrator position uncertainty is thought to be greater
  than the relative error introduced in phase-referencing.  Quoted
  proper motion and parallax uncertainties are most compact 68\%
  confidence intervals.  All astrometry is performed and reported in
  Equinox J2000.  Reported positions are for epoch 2002.0.  }
\end{deluxetable*}

\begin{deluxetable*}{lrrccrr} 
\tablecolumns{7}
\tablewidth{0pc}
\tablecaption{Parameters derived from astrometry\label{Tab:derivpars}}
\tablehead{
  \colhead{Pulsar} & \colhead{$l$} & \colhead{$b$} & \colhead{DM} &
  \colhead{$D_{DM}$}  & \colhead{$D_{\pi}$} & \colhead{$V_{\perp}$} \\ 
  \colhead{} & \colhead{(\arcdeg)} & \colhead{(\arcdeg)} &
  \colhead{(pc cm$^{-3}$)} & \colhead{(kpc)} & \colhead{(kpc)} &
  \colhead{(\kms)} 
}
\startdata 
B0031$-$07  & 110.42 & $-$69.82 & 11.38 & 0.4 & $ 1.06^{+0.08}_{-0.09} $ & $  77^{+6}_{-6}$  \\
  B0136+57  & 129.22 & $-$4.04  & 73.78 & 2.8 & $ 2.65^{+0.35}_{-0.26} $ & $ 319^{+42}_{-33}$ \\
  B0450$-$18  & 217.08 & $-$34.09 & 39.90 & 2.4 & $ 0.76^{+0.36}_{-0.60} $ & $  54^{+23}_{-45}$  \\
  B0450+55  & 152.62 & 7.55  	& 14.50 & 0.7 & $ 1.19^{+0.07}_{-0.06} $ & $ 317^{+19}_{-15}$  \\
J0538+2817  & 179.72 & $-$1.69  & 39.57 & 1.2 & $ 1.30^{+0.22}_{-0.16} $ & $ 357^{+59}_{-43}$  \\
B0818$-$13  & 235.89 & 12.59 	& 40.94 & 2.0 & $ 1.96^{+0.17}_{-0.12} $ & $ 418^{+36}_{-25}$  \\
  B1508+55  & 91.33  & 52.29    & 19.61 & 1.0 & $ 2.10^{+0.13}_{-0.14} $ & $ 963^{+61}_{-64}$  \\
  B1541+09  & 17.81  & 45.78    & 35.24 & $>$35 & $  7.2^{+1.3}_{-1.1} $   & $ 277^{+52}_{-42}$ \\
J1713+0747  & 28.75  & 25.22 	& 15.99 & 0.9 & $ 1.05^{+0.06}_{-0.07} $ & $  30^{+2}_{-2}$  \\
  B1933+16  & 52.44  & $-$2.09  & 158.52& 5.6 & $  5.2^{+1.5}_{-2.7} $   & $ 394^{+108}_{-208}$ \\
B2045$-$16  & 30.51  & $-$33.08 & 11.46 & 0.6 & $ 0.95^{+0.02}_{-0.02} $ & $ 512^{+13}_{-12}$  \\
  B2053+36  & 79.13  & $-$5.59  & 97.31 & 4.6 & $  5.5^{+1.2}_{-0.8} $   & $  69^{+15}_{-11}$  \\
  B2154+40  & 90.49  & $-$11.34 & 70.86 & 3.7 & $  3.4^{+0.9}_{-0.7} $   & $ 264^{+73}_{-52}$  \\
  B2310+42  & 104.41 & $-$16.42 & 17.28 & 1.2 & $ 1.06^{+0.08}_{-0.07}$  & $ 125^{+10}_{-8}$  \\
\enddata
\tablecomments{Pulsar distances $D_{DM}$ were estimated from their dispersion
  measure (DM) using the NE2001 electron density model \citep{cl02}.
  Confidence intervals for the parallax distances $D_\pi$ and
  transverse velocities $V_\perp$ were estimated directly from the
  bootstrap.}
\end{deluxetable*}

Compared to simple least squares fitting, the bootstrap method is
expected to return larger fit uncertainties, since the time sampling
for each trial in the bootstrap process is generally worse than simply
using all epochs of data in the fit process.  To compare the
performance of the bootstrap method to simple least squares fitting,
we simulated observations of PSR~J1713+0747 on the actual observation
dates, and recovered astrometric parameters both by a least squares
fit to all the data, and by bootstrap.  As expected, for Gaussian
random uncertainties in the input position at each epoch, the two
methods returned astrometric parameters that were essentially
identical, differing by much less than the fit uncertainties, and the
bootstrap method returned somewhat larger confidence intervals
compared to the simple least squares fit (between 0 and 25\%).  Next,
we added unmodeled systematic errors to the simulated positions, such
that the mean ``observed'' position at a given epoch differed from the
``true'' position and the observation uncertainties understated the
effective simulated errors.  As systematic errors were added to
increasing numbers of epochs, the results of the two techniques began
to diverge.  While the recovered best-fit astrometric parameters did
not differ significantly between the two methods, the bootstrap fit
uncertainty intervals generally encompassed the ``true'' parameter
values (within 1- or 2-$\sigma$) while the simple least squares method
returned uncertainties that were too optimistic and failed to include
the input values within several multiples of $\sigma$ with increasing
systematic errors.  These simulations therefore justify our decision
to use the bootstrap method when we are not able to reliably quantify
the systematic errors in our observations; while our astrometry may be
somewhat less {precise} as a result, it is more likely to be more
{accurate}. 

For illustration purposes, we show the parallax and proper motion
signature of PSR~B1933+16 in \Fref{sky}, and the parallax signature of
PSRs~B0818$-$13 and B2310+42 in \Fref{pi}.  In these figures, the
position error bars include estimates of the systematic error from the
scatter between positions measured at each epoch at different
frequency bands.  However, we note that these estimates are used only
for illustrative purposes, and the fit uncertainties are obtained from
bootstrap, as described above.

\begin{figure}[t!]
\epsscale{0.75}
\plotone{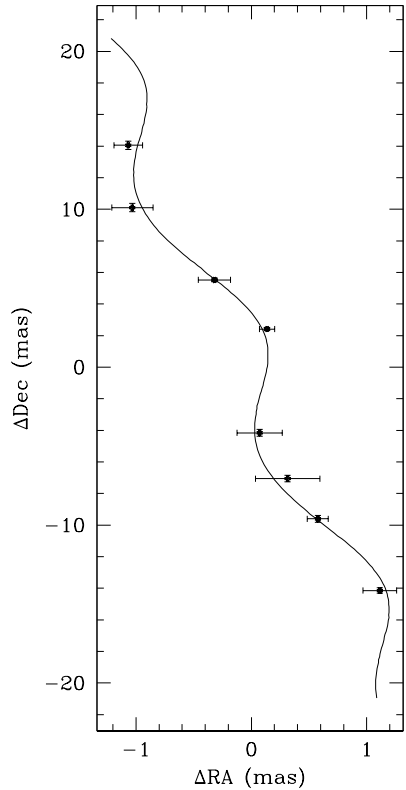}
\caption{The proper motion and parallax of PSR~B1933+16. The best fit
  parallax and proper motion ($\pmra = 1.1$~\mpy, $\pmdec =
  -16.1$~\mpy, $\pi = 0.2$~mas) is overplotted.  The earliest position
  (top left) is at MJD~52485, while the final position (bottom right)
  is at MJD~53116.  $\Delta$RA is the angular displacement
  $\Delta\alpha\, \cos\delta$. Note the different scales on the two
  axes.
\label{Fig:sky}
}
\end{figure}

\begin{figure*}[t!]
\epsscale{0.95}
\plotone{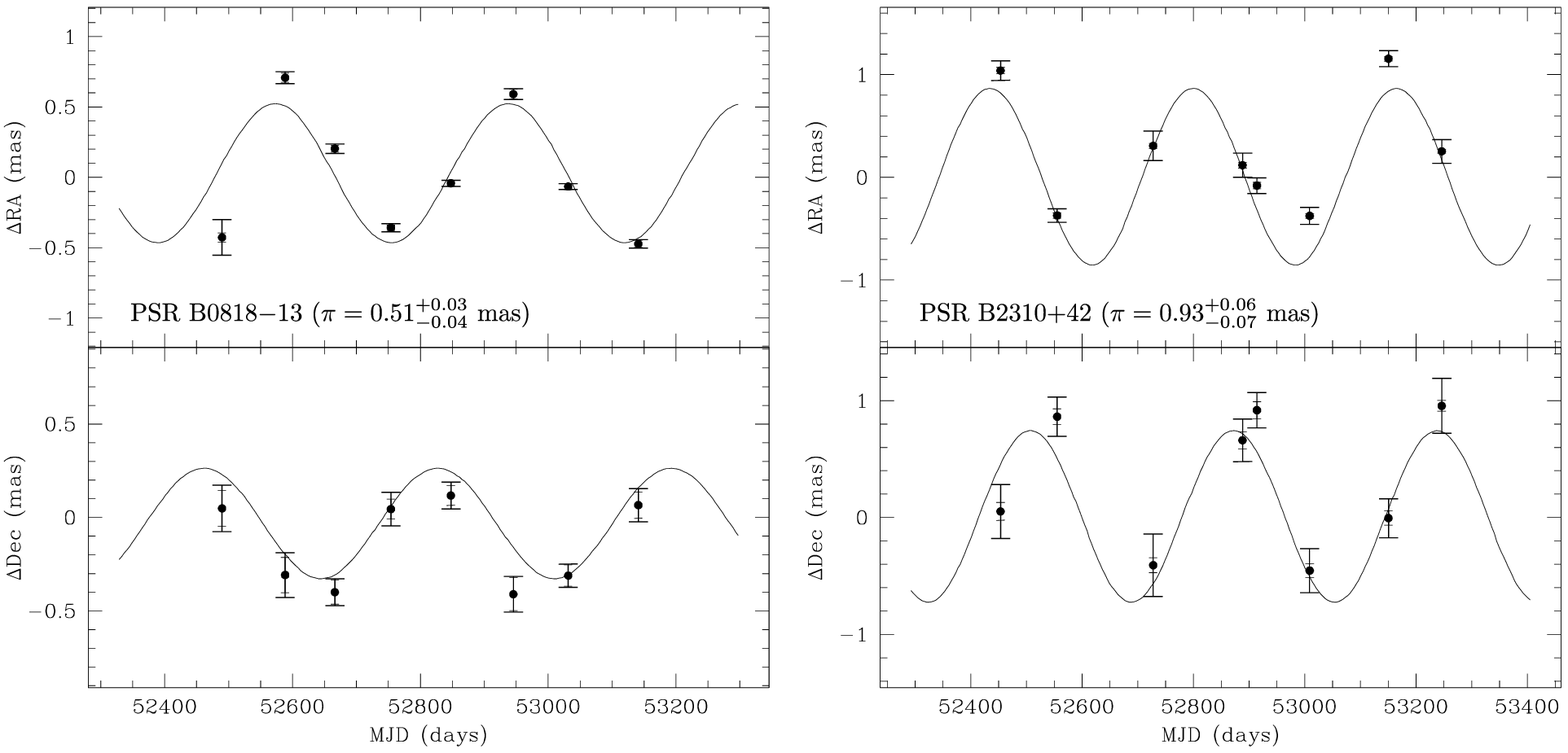}
\caption{The parallax signature of two pulsars in Right Ascension and
  Declination, after subtracting the respective best-fit proper
  motions from the astrometric positions. $\Delta$RA is the angular
  displacement $\Delta\alpha\, \cos\delta$.  Left: The parallax of
  PSR~B0818$-$13, with curves overplotted corresponding to the best
  fit parallax $\pi = 0.51$~mas. Right: The parallax of PSR~B2310+42,
  with curves overplotted corresponding to the best fit parallax $\pi
  = 0.93$~mas. In each panel, the inner error bars indicate the random
  position uncertainties, while the outer error bars indicate the net
  estimated uncertainties (random and systematic, added in quadrature)
  for illustration purposes only.
\label{Fig:pi}
}
\end{figure*}

Finally, we note that signal-to-noise considerations required the
averaging of all frequency bands for PSR~B0450$-$18.  As such, the
bootstrap was performed with only 8 position measurement pairs
($\alpha_t,\delta_t$ at each epoch) while fitting for 5 astrometric
parameters, resulting in a high proportion of degenerate samples.  For
this one object, we repeated the bootstrap for $10^6$ iterations, but
obtained no difference in the results or their confidence intervals
compared to $10^5$ iterations.  However, the uncertainties for the
astrometric parameters for PSR~B0450$-$18 are much larger than for the
rest of our sample.

\section{Discussion}\label{Sec:discuss}

We have presented parallaxes and proper motions for 14 pulsars
(\Tref{results}) and derived estimates for their distances and
transverse velocities (\Tref{derivpars}). These results represent a
significant increase in the number of pulsars with trigonometric
parallaxes, and include distances to two pulsars over 5~kpc away that
are determined to better than $\sim$15\% : most probable $D =
7.2^{+1.3}_{-1.1}$~kpc for PSR~B1541+09, and $5.5^{+1.2}_{-0.8}$~kpc
for PSR~B2053+36.  Early results from this program were published for
two objects in the sample, PSR~B1508+55 \citep{cvb+05} and
PSR~J0538+2817 \citep{nrb+07}, but with a simpler, more {\em ad hoc}
treatment of the systematic errors.  The bootstrap approach followed
here does not assume that errors at each epoch are normally
distributed, a condition that is now appreciated to be rarely true in
phase referenced astrometry at frequencies below 5~GHz, where the
turbulent ionosphere dominates the phase-referencing errors.  For
situations with over-determined measurements, such as those presented
here, the bootstrap method effectively results in discrepant data
receiving a lower weight, and so the current results are more robust.

For PSR~B1508+55, the astrometric parameters determined here are
consistent with our previous work, and the potential birth site in the
Cygnus OB associations is unchanged.  Nor is the high estimated
velocity ($V_\perp \approx 1000$~\kms; $V_{3d} \gtrsim 1100$~\kms)
significantly altered.  For PSR~J0538+2817, the proper motion in
declination received an adjustment of about 2~$\sigma$ when employing
the bootstrap analysis, but the magnitude of the change is small (from
$52.59 \pm 0.13$~\mpy\ to $52.87^{+0.09}_{-0.10}$~\mpy), and does not
affect the conclusions of \citet{nrb+07}.  The results also remain
consistent with the timing results of \citet{klh+03}, but are much
more precise.  Below we discuss some of the implications of our new
measurements, particularly for birth sites, kinematic ages, Galactic
electron density models, and reference frame ties.

\subsection{Pulsar Birth Sites and Kinematic Ages}\label{Sec:kin}

Accurate astrometric measurements allow pulsars to be traced back
through the Galaxy to potential birth sites
\citep[e.g.][]{hbz00,hbz01,vcc04}.  If a birth site can be identified,
the three-dimensional birth velocity can be estimated, a kinematic age
$\tau_{\rm kin}$ can be inferred, and a combination of the birth spin
period and braking index can be constrained.  A detailed analysis of
the Galactic trajectories of all pulsars with precise astrometry is
beyond the scope of this work and will be presented in future
(Vlemmings et al., in prep.), but a preliminary assessment of the
current sample of pulsars is presented here.  

The trajectories of the pulsars in our sample (except the old,
recycled pulsar~J1713+0747) are shown in \Fref{orbits} for three
different values of the (unknown) radial velocity ($V_{\rm r}=-200, 0$
and $+200$~\kms). The pulsar trajectories were calculated using the
three component St{\"a}ckel potential from \citet{fd03} with the
scaling parameters as described in \citet{vcc04}. \Fref{orbits} also
shows the positions and extent of the nearby OB associations from
\citet{zhb99} and a number of open clusters compiled from the WEBDA
catalog\footnote{http://www.univie.ac.at/webda/} with a distance
$<5$~kpc and a well-defined age $<10$~Myr. We only plot those open
clusters for which a pulsar trajectory crosses within $20$~pc of the
cluster center on the sky, without accounting for the distance or the
motion of the cluster itself.  The motion of the cluster can be $\sim$
several hundred pc during the lifetime of a pulsar, and needs to be
taken into account in a more detailed analysis.  Although a number of
pulsar trajectories pass close to the OB associations or open clusters
on the sky, the pulsar distance is typically incompatible with an
origin in the cluster or association. However, besides the earlier
identification of the likely birth site of PSR~B1508+55 in one of the
Cygnus OB associations \citep{cvb+05}, the coarse analysis performed
here reveals the possible birth location of PSR~B2045$-$16 in the open
cluster NGC~6604, as discussed below.

\begin{figure}[t!]
\epsscale{1.1}
\plotone{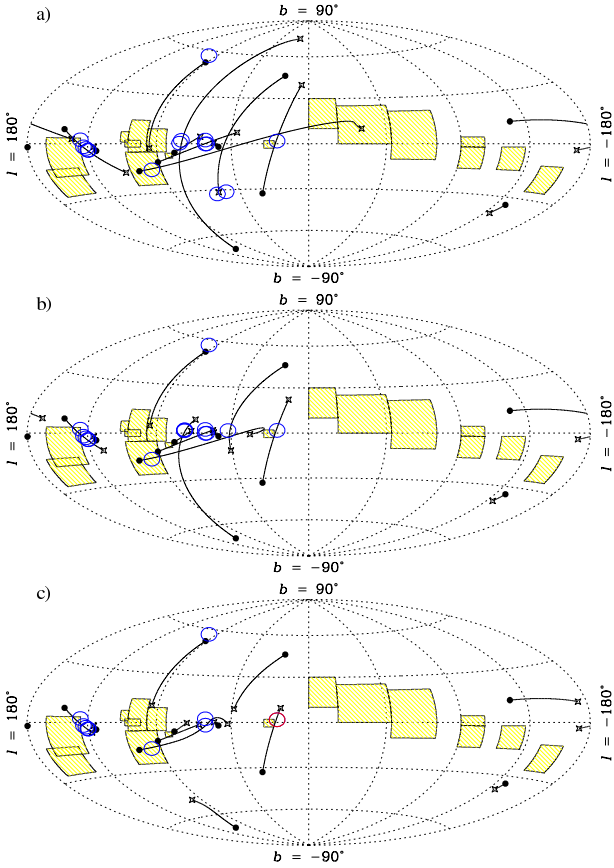}
\caption{The distribution of our pulsar sample in Galactic
  coordinates. The solid dots denote the present pulsar positions, and
  the lines represent their trajectory on the sky calculated for an
  age corresponding to the pulsar spin-down age. The open stars
  indicate the possible birth places of the pulsars. The three panels
  {\it a,b} and {\it c} represent the calculated orbits using
  $V_r=+200, 0$ and $-200$~\kms\ respectively. From left to right
  (decreasing $l$ of the present position) the pulsars are J0538+2817,
  B0450+55, B0136+57, B0031$-$07, B2310+42, B1508+55, B2154+40,
  B2053+36, B1933+16, B2045$-$16, B1541+09, B0818$-$13 and
  B0450$-$18. The shaded areas indicate the nearby OB associations
  \citep{zhb99}. From left to right and from top to bottom: Per~OB2,
  $\alpha$~Per, Cep~OB3, Cep~OB6, Cep~OB2, Lac~OB1, Cyg~OB7, Cyg~OB4,
  Sct~OB2, Upper Sco, Upper Cen Lupus, Lower Cen Crux, Tr~10,
  Vela~OB2, Col~121 and Ori~OB1. The open circles denote the open
  clusters from the WEBDA catalog with well-defined distances
  ($<5$~kpc), positions and ages ($<50$~Myr). Only those open clusters
  are shown for which a pulsar orbit approaches the cluster position
  within $<20$~pc {\em on the sky}. \label{Fig:orbits} }
\end{figure}

The progenitors of NS are massive O- and B-stars, which have a low
Galactic scale height $\sim$63~pc, much smaller than the
characteristic distance of radio pulsars from the Galactic plane.
This is consistent with pulsars having significantly larger peculiar
velocities than their progenitors, as recognized for example by
\citet{ggo70}.
As such, it is expected that most (young) pulsars should be moving
away from the Galactic plane.  Although \citet{hla93} found that 17\%
of the young pulsars in their sample seemed to be moving towards the
plane, later analysis indicates that after accounting for the pulsar
birth scale height \citep[$z_0 = 160 \pm 40$~pc,][]{acc02}, all young
pulsars do seem to be moving away from the Galactic plane
\citep[e.g.,][]{cc98, bfg+03, hllk05}.  Any apparent motion toward the
Galactic plane can often be explained by the unknown radial velocity
$V_{\rm r}$, as the velocity perpendicular to the plane $V_{\rm z}$ is
given by
\begin{equation}
V_{\rm z}=V_{\rm b}\cos{b}+V_{\rm r}\sin{b},
\end{equation} 
where $V_{\rm b}$ is the component of the velocity in the tangent
plane along Galactic latitude $b$.

Alternatively, pulsars older than $\sim20$~Myr could potentially
already be falling back towards the plane due to the effect of the
Galactic potential on their trajectory.  \Fref{vzz} indicates $V_{\rm
  z}$ against the location with respect to the Galactic plane for the
pulsars in our sample.  It is apparent that for a reasonable range of
$V_{\rm r}$, all the pulsars are moving away from the Galactic plane,
with the exception of PSR~J0538+2817, which is located well within the
pulsar birth scale height at $z=-41$~pc.  \Fref{vzz} also shows that,
as expected, the youngest pulsars are found closest to the plane.

Assuming that all pulsars are indeed born in the Galactic plane
between $-160<z_0<160$~pc, we can estimate the age of the pulsar from
the measured $z = D \sin{b}$ and $V_{\rm z}$.  The kinematic age
$\tau_{\rm kin}$ is given by:
\begin{equation}
\tau_{\rm kin}=(D\sin{b}-z_0)/V_{\rm z}.
\end{equation}
The estimated kinematic ages of the pulsars in our sample are shown in
\Fref{tkin}, for $-200<V_{\rm r}<+200$~\kms.  As deceleration in the
Galactic potential will reduce $V_{\rm z}$ as pulsars get older,
$\tau_{\rm kin}$ will be a lower limit on the true age.  Spin-down
ages $\tau_{\rm sd} = P/2\dot{P}$ are also shown for comparison in
\Fref{tkin}, where $P$ is the period of the pulsar, assumed to have
been born spinning rapidly and slowing steadily with a braking index
$n=3$.  Each of these assumptions (rapid spin at birth, steady
slowdown, $n=3$) is known to be problematic in specific cases
\citep[e.g.,][]{gf00, klh+03}.  However, in spite of the variety of
assumptions built into both $\tau_{\rm sd}$ and $\tau_{\rm kin}$, the
agreement between them for most of the pulsars in our sample is
remarkable.

\begin{figure}[t!]
\epsscale{1.0}
\plotone{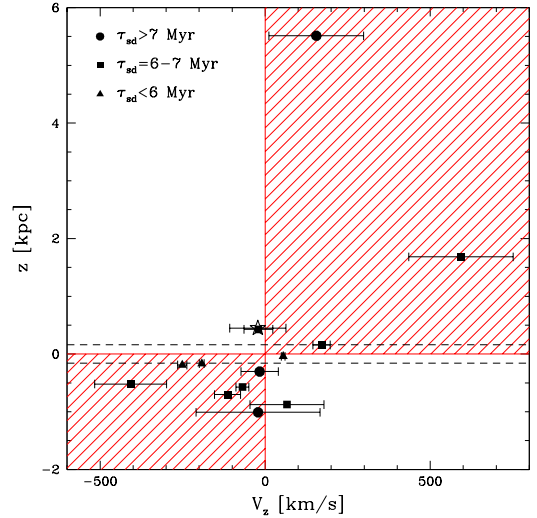}
\caption{Pulsar distance above the Galactic plane ($z$) versus its
  velocity perpendicular to the plane ($V_{\rm z}$). The different
  symbols represent different ranges of pulsar spin-down age and the
  open star indicates the recycled pulsar J1713+0747. As $V_{\rm z}$
  depends on the unknown radial velocity of the pulsar, the bars
  denote $-200<V_{\rm r}<+200$~\kms. The pulsars in the shaded areas
  are moving away from the Galactic plane while those in the open
  areas are falling towards the plane. The horizontal dashed lines
  indicate the pulsar birth scale height determined by
  \citet{acc02}.\label{Fig:vzz} }
\end{figure}

\begin{figure}[t!]
\epsscale{1.0}
\plotone{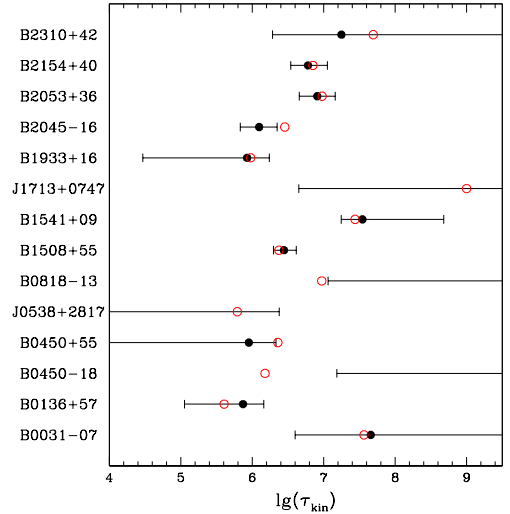}
\caption{Kinematic ages for our pulsar sample. The solid symbols
  denote the kinematic ages calculated using the pulsar distance above
  the Galactic plane and velocity $V_{\rm z}$, assuming $V_{\rm
    r}=0$~\kms. The bars indicate the range $-200<V_{\rm
    r}<+200$~\kms. The open circles are the spin-down ages. For those
  pulsars without solid symbols, the perpendicular velocity is
  pointing towards the plane when $V_{\rm r}=0$~\kms. As deceleration
  in the Galactic potential will have affected $V_{\rm z}$ for the
  pulsars older than $\sim 10$~Myr, $\tau_{\rm kin}$ will be a lower
  limit.\label{Fig:tkin} }
\end{figure}

The only pulsars for which there is an apparent discrepancy between
the two age estimates are PSRs~J1713+0747, B0818$-$13, J0538+2817 and
B0450$-$18.  These pulsars appear to be falling towards the Galactic
plane and thus have undetermined $\tau_{\rm kin}$ for $V_{\rm r} = 0$.
The recycled pulsar J1713+0747 has likely already completed one or
more orbits through the Galactic plane, and the simple analysis
outlined above is obviously inapplicable in that case.  Allowing a
radial velocity slightly larger than the arbitrary cutoff of
200~\kms\ resolves the discrepancy for PSR~B0818$-$13. PSR~J0538+2817
is still within the Galactic pulsar birth scale height, as indicated
above.  It is associated with the known supernova remnant S147, and
the pulsar age of $\sim 30$~kyr derived from the association
\citep{klh+03} is also quite inconsistent with its spin-down age
($\tau_{\rm sd}=620$~kyr).  PSR~B0450$-$18 requires $V_r \gtrsim
360$~\kms\ at its most probable distance inferred from bootstrap ($D =
0.76$~kpc), which may not be unreasonable, but in any case it has the
largest uncertainties among the objects in our sample
(c.f.~\Sref{analysis}).

\subsubsection{PSR B2045$-$16}

In our analysis of pulsar trajectories, PSR~B2045$-$16 was found to
pass on the sky within $\sim20$~pc of the open cluster NGC~6604 for a
radial velocity $V_{\rm r}\approx -200$~\kms.  Since the distance
between the cluster and the pulsar was also $\lesssim 20$~pc, NGC~6604
could be the birth location of PSR~B2045$-$16.  If it was born in this
cluster, the kinematic age of PSR~B2045$-$16 is $\tau_{\rm kin}
\approx 1.9$~Myr, slightly less than its spin-down age $\tau_{\rm sd}
= 2.85$~Myr.  NGC~6604 is an open cluster with an age $\sim 6.5$~Myr
\citep{kpr+05}.  We infer that the progenitor of PSR~B2045$-$16 was
probably more massive than $\sim20$~M$_\odot$ and that the birth
velocity of the pulsar was $\sim575$~\kms.  Furthermore, the mismatch
between the spin-down age and kinematic age can be resolved if the
initial spin period of the pulsar was $P_0 \sim 360$~ms, assuming a
fixed braking index $n=3$, consistent with the population birth spin
period distribution ($300 \pm 150$~ms) found by \citet{fk06}.
Alternatively, if the initial spin period is assumed to be much
smaller than its current period, the kinematic age implies a braking
index $n \approx 4$, which is larger than all measured values to date
and thus an unlikely explanation.  Further analysis will need to
confirm the association of PSR~B2045$-$16 with NGC~6604, taking into
account the proper motion of NGC~6604 \citep[$\mu_\alpha =
  -0.36$~\mpy, $\mu_\delta = -2.68$~\mpy][]{kpr+05}, which corresponds
to a motion on the sky of $\sim60$~pc in 2~Myr.

\subsection{Electron Density Models}\label{Sec:ne}

Model independent distances to pulsars provide essential calibration
points for models of the Galactic electron density distribution such
as NE2001 \citep{cl02}, and one of the objectives of our observations
is to use new pulsar distances to improve successors to NE2001.  We
have compared our astrometric distance estimates with those obtained
from NE2001, as illustrated in \Fref{ne}.  In constructing NE2001,
\citet{cl02} estimated the uncertainty in the distances derived from
the pulse dispersion measure (DM) to be 20\%.  For the purposes of our
comparison, we do not include the NE2001 uncertainty estimate, but
consider only whether the NE2001 distance estimate is within the 68\%
and 99\% confidence intervals of the parallax distance.

\begin{figure}[t!]
\epsscale{1.2}
\plotone{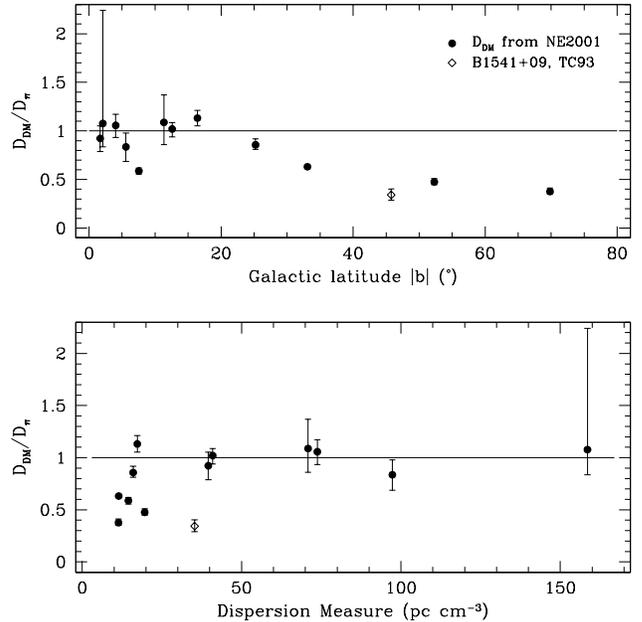}
\caption{Comparison of distance estimates from dispersion measure and
  parallax distances for the objects in our sample as a function of
  (top panel) absolute Galactic latitude $|b|$ and (bottom panel)
  dispersion measure $\int_0^D n_e\, dl$.  Dispersion measure
  distances are from NE2001 \citep{cl02} except for PSR~B1541+09, for
  which NE2001 provides only a lower limit on the distance (see
  discussion in \Sref{ne}), and we have used an earlier distance
  estimate from \citet{tc93}.  Error bars reflect only the
  uncertainties in the distance estimated from parallax, and
  PSR~B0450$-$18 has been excluded from the plot due to the large
  uncertainties.  It is apparent that NE2001 underestimates distances
  to some of the pulsars in our sample at high Galactic latitudes,
  consistent with a larger scale height or irregularities in the ISM
  at the relevant scales. \label{Fig:ne} }
\end{figure}

Of the 14 pulsars in our sample, the NE2001 distance estimates are in
approximate agreement with our parallax determinations (within or just
beyond the 68\% confidence intervals) for~9 of the pulsars.  The other
five pulsars (B0031$-$07, B1508+55, B2045$-$16, B1541+09, and
B0450+55) are discussed below.   

Three of the discrepant pulsars (\objectname[PSR]{B0031$-$07},
\objectname[PSR]{B1508$+$55}, and \objectname[PSR]{B2045$-$16}) have
high Galactic latitudes $|b| > 30\arcdeg$ and low DMs $\mathrm{DM} <
20$~pc~cm$^{-3}$.  The NE2001 model contains a set of low electron
density components designed to reproduce the effects of local Galactic
structure, namely regions such as the Local Hot Bubble (LHB), a local
Low Density Region (LDR), and the Local Superbubble (LSB).  The
distances to these three pulsars are large enough that none are within
any of these local, low-density regions.  Thus, these local structures
cannot be the explanation for the systematic discrepancies.

\cite{gmcm08} have re-analyzed the available DM data toward high
Galactic latitude pulsars and find that the scale height of the
ionized gas is approximately 1.8~kpc, about a factor of~2
higher than the thick disk component in the NE2001 model.  Analyzing
the DM data in combination with emission measures (EM) derived from
H$\alpha$ observations, they also conclude that the filling factor of
the ionized gas increases with distance above the Galactic plane.

Using the \cite{gmcm08} re-assessment for the ionized gas scale
height, we have estimated the distances to these pulsars from their
DMs.  In general, the resulting distances are slight {over-estimates}
relative to their parallax distances.  We conclude that the new
parallax distances and the DMs of these pulsars are consistent with
the conclusion of \cite{gmcm08} of a larger scale height for the
ionized gas, though their estimated scale height of 1.8~kpc may be a
bit too high.  A revised fit including the new distances presented
here indeed yields a slightly lower scale height (B.~M.~Gaensler,
personal communication), but does not significantly alter the
conclusions of \citet{gmcm08}.  We note, however, that the ISM is
likely to be significantly irregular or patchy on the relevant length
scales, particularly at high Galactic latitudes, where chimmneys and
voids are important contributors to the structure. 

\objectname[PSR]{PSR~B1541$+$09} is also at high Galactic latitude ($b
= 45\arcdeg$) but has a relatively large DM (35.24~pc~cm${}^{-3}$).
This DM is above the maximal value ($\approx 30$~pc~cm$^{-3}$) from
the thick disk component in the NE2001 model, 
which only provides lower limits on the pulsar distance.  Thus, in
order to reproduce the DM, the NE2001 model includes a ``clump'' of
enhanced electron density (contributing about~10\% of the total DM),
though \cite{cl02} could find no obvious feature along the line of
sight that might generate this clump (e.g., an \ion{H}{2} region or~O
star).  This line of sight also passes above two spiral arms in the
NE2001 model.  One possibility is that the scale height of the ionized
gas above one or both of these arms is larger than included in the
model.

Lastly, \objectname[PSR]{PSR~B0450$+$55} lies at a relatively low
Galactic latitude ($b = 7\fdg5$), with DM $= 14.495$~pc~cm$^{-3}$.  At
the parallax distance to the pulsar, the DM inferred from the NE2001
model is 33~pc~cm${}^{-3}$, much larger than what is observed.  The
situation can plausibly be explained by a deficit of electrons in this
direction.  Within the NE2001 model, there are attempts to account for
such deficits (``voids'') in the electron density along specific lines
of sight, but such modeling is limited by the sparseness of available
DM-independent distance estimates.

We have examined both the Virginia Tech Spectral-Line Survey
H-$\alpha$ image and the WHAM H-$\alpha$ survey \citep{hrt+03};
neither shows any indication that the line of sight toward the pulsar
has any deficit of H-$\alpha$ emission compared to nearby lines of
sight.  However, a CO survey \citep{dht01} shows a tongue of CO
emission that crosses the line of sight to the pulsar.  While the
filling factor of the CO gas along the line of sight to the pulsar may
not be large, its presence suggests that not all of the line of sight
may be ionized, consistent with an apparent void in the electron
density along the line of sight to this pulsar.

\subsection{Reference Frame Ties}\label{Sec:frametie}

Pulse timing provides radio pulsar positions in the Solar system
reference frame, while VLBI measurements are tied to the distant
quasars.  Simply measuring precise positions for pulsars via the two
different techniques enables fundamental reference frame ties between
the Solar system and the extragalactic ICRF \citep[e.g.][]{bcr+96}.
The recycled pulsar J1713+0747 is now one of two pulsars \citep[along
  with PSR~J0437$-$4715;][]{dvtb08} with high-precision astrometry and
statistically significant measurements of parallax using {\em both}
pulse timing and VLBI.  In \Tref{J1713} we list the astrometric
parameters for PSR~J1713+0747 as determined by interferometry (this
work) as well as by two independent pulse timing efforts
\citep{sns+05,hbo06}.


\begin{deluxetable}{lrrr} 
\tablecolumns{4}
\tablewidth{0pc}
\tablecaption{Astrometry for PSR~J1713+0747\label{Tab:J1713}}
\tablehead{ \colhead{Measurement} & \colhead{VLBA} & \colhead{Splaver
    et al.} & \colhead{Hotan et al.} } 
\startdata 
$\Delta\alpha$ (s)&
	0.5306(1) &
	0.5307826(7) &
	0.53077(1)\\
$\Delta\delta$ (\arcsec)    & 
	0.519(2)  &
	0.52339(3) &
	0.5228(2)  \\
$\mu_{\alpha}$ (mas yr$^{-1}$) &  
	4.75$^{+17}_{-7}$ &
	4.917(4) &
	4.97(6) \\
$\mu_{\delta}$  (mas yr$^{-1}$) & 
	$-$3.67(16) &
	$-$3.933(10) &
	$-$3.7(1) \\
$\pi$ (mas) & 
	0.95(6) &
	0.89(8) &
	1.1(1) \\
\enddata
\tablecomments{Astrometric parameters for PSR~J1713+0747 from VLBA
  astrometry (this work) are compared to parameters derived
  independently from pulse timing \citep{sns+05,hbo06}.  All
  parameters are for equinox J2000; positions are shifted to match the
  VLBI data epoch of year 2002.0 (MJD 52275) and listed as offsets
  from RA~$17\rah 13\ram 49\ras$ ($\Delta\alpha$),
  Dec~$07\dd 37\am 47\as$ ($\Delta\delta$).  }
\end{deluxetable}

The data used by \citet{sns+05} span 12 years (though with a 4 year
gap), allowing compensation for timing noise.  In contrast,
\citet{hbo06} used data spanning only 2.5 years, which did not allow
corrections for timing noise, but allowed them to minimize systematics
through the consistency of their instrumental setup.  We note that the
overall consistency between the timing and VLBI results for proper
motion and parallax is quite good, although the position differs in
Declination at the $\sim$mas level.  The longer time series allows
\citet{sns+05} to attain higher precision in their proper motion
estimates, but the uncertainties in parallax are comparable between
the 12 year timing data set and the 2 year VLBI data, illustrating the
complementarity of the techniques.  An ensemble of such measurements
on recycled pulsars has great promise for reference frame ties, as
well as for the detection of gravitational waves.

\section{Conclusions}\label{Sec:final}

By simply measuring the positions of neutron stars over time to high
accuracy, it is possible to establish constraints on a variety of
scientific questions.  Here we have presented the motivation, methods,
and results from a large VLBA astrometry program.  We have described
our initial VLA surveys to select targets and find in-beam
calibrators, as well as VLBA observations, pulsar gating, and data
reduction.  We have identified a few specific pitfalls, including
proximity of the targets to the Sun, and the widespread presence of
satellite-generated RFI.  We have also described the use of the
bootstrap method to comprehensively treat unmodeled systematic errors.
The methods described here are serving as templates for our ongoing
astrometry programs, and may serve as useful guidelines for the wider
community. 

As a result of our VLBA astrometry program, we have measured new
parallaxes and proper motions for 14 pulsars, including $\pi =
0.13^{+0.02}_{-0.02}$~mas for PSR~B1541+09.  The measurements have
been exploited to investigate their kinematics and birth sites.  Young
pulsars are found to be leaving the Galactic plane, as expected, and
we also find that spindown ages and kinematic ages are generally in
reasonable agreement, with some notable exceptions.  We have
identified a plausible birth site for PSR~B1508+55 in one of the Cyg
OB associations \citep{cvb+05} and for PSR~B2045$-$16 in the open
cluster NGC~6604 (this work).

The new model-independent distances have also allowed us to revisit
models of the Galactic electron density distribution.  Comparing
distance estimates from parallax and pulse dispersion measure, we find
that NE2001 underestimates distances for some objects at high Galactic
latitudes, while for others the estimates agree with or are larger
than the parallax distances.  Our findings are consistent with a
larger scale height or with irregularities in the ISM on relevant
length scales.  Independent distance measurements are essential to
calibrate models and probe structure in the ISM, and the present
sample will be incorporated into future iterations of the electron
density distribution model, ultimately improving DM-based distance
estimates for all pulsars.

Finally, we have compared our precise astrometry on the recycled
pulsar J1713+0747 with two independent results from pulse timing.  The
comparison can be used to verify consistency between the extragalactic
ICRF and the Solar system reference frame, especially as part of an
ensemble of recycled pulsars. 

The Very Long Baseline Array is the only instrument that offers
full-time, dedicated VLBI capabilities.  Our results once again
illustrate the power of the instrument to provide astrometry even at
lower frequencies with high precision and excellent accuracy.

\acknowledgements{ We acknowledge the Very Long Baseline Array
  operations team for their efforts in scheduling and supporting our
  large VLBA astrometry program. The National Radio Astronomy
  Observatory (NRAO) is a facility of the National Science Foundation
  (NSF) operated under cooperative agreement by Associated
  Universities, Inc.  SC thanks David Kaplan for useful discussions
  about astrometric errors,
  and Bryan Gaensler for useful discussions about electron density
  models.  We also thank the anonymous referee for carefully reading the
  manuscript and providing valuable feedback, and Adam Deller for
  drawing our attention to a transcription error in the manuscript.
  SC acknowledges support from the University of Sydney
  Postdoctoral Fellowship program, and he was a Jansky Fellow of the
  NRAO at the time this large project was initiated.  Basic research
  in radio astronomy at the NRL is supported by 6.1 Base funding.  SET
  acknowledges support from the NSF (grant AST 0506453).  This
  research has made use of the WEBDA database, operated at the
  Institute for Astronomy of the University of Vienna.  The Wisconsin
  H-Alpha Mapper is funded by the NSF.  {\it Facilities:} {VLBA~()} 
}


\begin{thebibliography}{72}
\expandafter\ifx\csname natexlab\endcsname\relax\def\natexlab#1{#1}\fi

\bibitem[{{Arras} \& {Lai}(1999)}]{al99}
{Arras}, P., \& {Lai}, D. 1999, \apj, 519, 745

\bibitem[{{Arzoumanian} {et~al.}(2002){Arzoumanian}, {Chernoff}, \&
  {Cordes}}]{acc02}
{Arzoumanian}, Z., {Chernoff}, D.~F., \& {Cordes}, J.~M. 2002, \apj, 568, 289

\bibitem[{{Bartel} {et~al.}(1996){Bartel}, {Chandler}, {Ratner}, {Shapiro},
  {Pan}, \& {Cappallo}}]{bcr+96}
{Bartel}, N., {Chandler}, J.~F., {Ratner}, M.~I., {Shapiro}, I.~L., {Pan}, R.,
  \& {Cappallo}, R.~J. 1996, \aj, 112, 1690

\bibitem[{{Beasley} {et~al.}(2002){Beasley}, {Gordon}, {Peck}, {Petrov},
  {MacMillan}, {Fomalont}, \& {Ma}}]{vcs1}
{Beasley}, A.~J., {Gordon}, D., {Peck}, A.~B., {Petrov}, L., {MacMillan},
  D.~S., {Fomalont}, E.~B., \& {Ma}, C. 2002, \apjs, 141, 13

\bibitem[{{Blazek} {et~al.}(2006){Blazek}, {Gaensler}, {Chatterjee}, {van der
  Swaluw}, {Camilo}, \& {Stappers}}]{bgc+06}
{Blazek}, J.~A., {Gaensler}, B.~M., {Chatterjee}, S., {van der Swaluw}, E.,
  {Camilo}, F., \& {Stappers}, B.~W. 2006, \apj, 652, 1523

\bibitem[{{Bridle} \& {Schwab}(1999)}]{bs99}
{Bridle}, A.~H., \& {Schwab}, F.~R. 1999, in Astronomical Society of the
  Pacific Conference Series, Vol. 180, Synthesis Imaging in Radio Astronomy II,
  ed. G.~B. {Taylor}, C.~L. {Carilli}, \& R.~A. {Perley}, 371--+

\bibitem[{{Briggs} {et~al.}(1999){Briggs}, {Schwab}, \& {Sramek}}]{bss99}
{Briggs}, D.~S., {Schwab}, F.~R., \& {Sramek}, R.~A. 1999, in Astronomical
  Society of the Pacific Conference Series, Vol. 180, Synthesis Imaging in
  Radio Astronomy II, ed. G.~B. {Taylor}, C.~L. {Carilli}, \& R.~A. {Perley},
  127--+

\bibitem[{{Brisken} {et~al.}(2000){Brisken}, {Benson}, {Beasley}, {Fomalont},
  {Goss}, \& {Thorsett}}]{bbb+00}
{Brisken}, W.~F., {Benson}, J.~M., {Beasley}, A.~J., {Fomalont}, E.~B., {Goss},
  W.~M., \& {Thorsett}, S.~E. 2000, \apj, 541, 959

\bibitem[{{Brisken} {et~al.}(2002){Brisken}, {Benson}, {Goss}, \&
  {Thorsett}}]{bbgt02}
{Brisken}, W.~F., {Benson}, J.~M., {Goss}, W.~M., \& {Thorsett}, S.~E. 2002,
  \apj, 571, 906

\bibitem[{{Brisken} {et~al.}(2003{\natexlab{a}}){Brisken}, {Fruchter}, {Goss},
  {Herrnstein}, \& {Thorsett}}]{bfg+03}
{Brisken}, W.~F., {Fruchter}, A.~S., {Goss}, W.~M., {Herrnstein}, R.~M., \&
  {Thorsett}, S.~E. 2003{\natexlab{a}}, \aj, 126, 3090

\bibitem[{{Brisken} {et~al.}(2003{\natexlab{b}}){Brisken}, {Thorsett},
  {Golden}, \& {Goss}}]{btgg03}
{Brisken}, W.~F., {Thorsett}, S.~E., {Golden}, A., \& {Goss}, W.~M.
  2003{\natexlab{b}}, \apjl, 593, L89

\bibitem[{{Burrows} \& {Hayes}(1996)}]{bh96}
{Burrows}, A., \& {Hayes}, J. 1996, Physical Review Letters, 76, 352

\bibitem[{{Chatterjee} {et~al.}(2001){Chatterjee}, {Cordes}, {Lazio}, {Goss},
  {Fomalont}, \& {Benson}}]{ccl+01}
{Chatterjee}, S., {Cordes}, J.~M., {Lazio}, T.~J.~W., {Goss}, W.~M.,
  {Fomalont}, E.~B., \& {Benson}, J.~M. 2001, \apj, 550, 287

\bibitem[{{Chatterjee} {et~al.}(2004){Chatterjee}, {Cordes}, {Vlemmings},
  {Arzoumanian}, {Goss}, \& {Lazio}}]{ccv+04}
{Chatterjee}, S., {Cordes}, J.~M., {Vlemmings}, W.~H.~T., {Arzoumanian}, Z.,
  {Goss}, W.~M., \& {Lazio}, T.~J.~W. 2004, \apj, 604, 339

\bibitem[{{Chatterjee} {et~al.}(2005){Chatterjee}, {Vlemmings}, {Brisken},
  {Lazio}, {Cordes}, {Goss}, {Thorsett}, {Fomalont}, {Lyne}, \&
  {Kramer}}]{cvb+05}
{Chatterjee}, S., {Vlemmings}, W.~H.~T., {Brisken}, W.~F., {Lazio}, T.~J.~W.,
  {Cordes}, J.~M., {Goss}, W.~M., {Thorsett}, S.~E., {Fomalont}, E.~B., {Lyne},
  A.~G., \& {Kramer}, M. 2005, \apjl, 630, L61

\bibitem[{{Cordes} \& {Chernoff}(1998)}]{cc98}
{Cordes}, J.~M., \& {Chernoff}, D.~F. 1998, \apj, 505, 315

\bibitem[{{Cordes} \& {Lazio}(2002)}]{cl02}
{Cordes}, J.~M., \& {Lazio}, T.~J.~W. 2002, ArXiv e-print, astro-ph/0207156

\bibitem[{{Cornwell} \& {Perley}(1992)}]{cp92}
{Cornwell}, T.~J., \& {Perley}, R.~A. 1992, \aap, 261, 353

\bibitem[{{Dame} {et~al.}(2001){Dame}, {Hartmann}, \& {Thaddeus}}]{dht01}
{Dame}, T.~M., {Hartmann}, D., \& {Thaddeus}, P. 2001, \apj, 547, 792

\bibitem[{{De Luca} {et~al.}(2008){De Luca}, {Caraveo}, {Esposito}, \&
  {Hurley}}]{dceh08}
{De Luca}, A., {Caraveo}, P.~A., {Esposito}, P., \& {Hurley}, K. 2008, ArXiv
  e-prints, 0810.3804

\bibitem[{{de Zeeuw} {et~al.}(1999){de Zeeuw}, {Hoogerwerf}, {de Bruijne},
  {Brown}, \& {Blaauw}}]{zhb99}
{de Zeeuw}, P.~T., {Hoogerwerf}, R., {de Bruijne}, J.~H.~J., {Brown}, A.~G.~A.,
  \& {Blaauw}, A. 1999, \aj, 117, 354

\bibitem[{{Deller} {et~al.}(2008){Deller}, {Verbiest}, {Tingay}, \&
  {Bailes}}]{dvtb08}
{Deller}, A.~T., {Verbiest}, J.~P.~W., {Tingay}, S.~J., \& {Bailes}, M. 2008,
  \apjl, 685, L67

\bibitem[{{Efron} \& {Tibshirani}(1991)}]{et91}
{Efron}, B., \& {Tibshirani}, R. 1991, Science, 253, 390

\bibitem[{{Famaey} \& {Dejonghe}(2003)}]{fd03}
{Famaey}, B., \& {Dejonghe}, H. 2003, \mnras, 340, 752

\bibitem[{{Faucher-Gigu{\`e}re} \& {Kaspi}(2006)}]{fk06}
{Faucher-Gigu{\`e}re}, C.-A., \& {Kaspi}, V.~M. 2006, \apj, 643, 332

\bibitem[{{Fey} {et~al.}(2004){Fey}, {Ma}, {Arias}, {Charlot},
  {Feissel-Vernier}, {Gontier}, {Jacobs}, {Li}, \& {MacMillan}}]{fma+04}
{Fey}, A.~L., {Ma}, C., {Arias}, E.~F., {Charlot}, P., {Feissel-Vernier}, M.,
  {Gontier}, A.-M., {Jacobs}, C.~S., {Li}, J., \& {MacMillan}, D.~S. 2004, \aj,
  127, 3587

\bibitem[{{Fomalont} {et~al.}(1999){Fomalont}, {Goss}, {Beasley}, \&
  {Chatterjee}}]{fgbc99}
{Fomalont}, E.~B., {Goss}, W.~M., {Beasley}, A.~J., \& {Chatterjee}, S. 1999,
  \aj, 117, 3025

\bibitem[{{Fomalont} {et~al.}(2006){Fomalont}, {Kellermann}, {Cowie}, {Capak},
  {Barger}, {Partridge}, {Windhorst}, \& {Richards}}]{fkc+06}
{Fomalont}, E.~B., {Kellermann}, K.~I., {Cowie}, L.~L., {Capak}, P., {Barger},
  A.~J., {Partridge}, R.~B., {Windhorst}, R.~A., \& {Richards}, E.~A. 2006,
  \apjs, 167, 103

\bibitem[{{Fomalont} {et~al.}(2003){Fomalont}, {Petrov}, {MacMillan}, {Gordon},
  \& {Ma}}]{vcs2}
{Fomalont}, E.~B., {Petrov}, L., {MacMillan}, D.~S., {Gordon}, D., \& {Ma}, C.
  2003, \aj, 126, 2562

\bibitem[{{Gaensler} \& {Frail}(2000)}]{gf00}
{Gaensler}, B.~M., \& {Frail}, D.~A. 2000, \nat, 406, 158

\bibitem[{{Gaensler} {et~al.}(2008){Gaensler}, {Madsen}, {Chatterjee}, \&
  {Mao}}]{gmcm08}
{Gaensler}, B.~M., {Madsen}, G.~J., {Chatterjee}, S., \& {Mao}, S.~A. 2008,
  ArXiv e-prints

\bibitem[{{Gott} {et~al.}(1970){Gott}, {Gunn}, \& {Ostriker}}]{ggo70}
{Gott}, J.~R.~I., {Gunn}, J.~E., \& {Ostriker}, J.~P. 1970, \apjl, 160, L91

\bibitem[{{Gwinn} {et~al.}(1986){Gwinn}, {Taylor}, {Weisberg}, \&
  {Rawley}}]{gtwr86}
{Gwinn}, C.~R., {Taylor}, J.~H., {Weisberg}, J.~M., \& {Rawley}, L.~A. 1986,
  \aj, 91, 338

\bibitem[{{Haffner} {et~al.}(2003){Haffner}, {Reynolds}, {Tufte}, {Madsen},
  {Jaehnig}, \& {Percival}}]{hrt+03}
{Haffner}, L.~M., {Reynolds}, R.~J., {Tufte}, S.~L., {Madsen}, G.~J.,
  {Jaehnig}, K.~P., \& {Percival}, J.~W. 2003, \apjs, 149, 405

\bibitem[{{Harrison} {et~al.}(1993){Harrison}, {Lyne}, \& {Anderson}}]{hla93}
{Harrison}, P.~A., {Lyne}, A.~G., \& {Anderson}, B. 1993, \mnras, 261, 113

\bibitem[{{Helfand} {et~al.}(2007){Helfand}, {Chatterjee}, {Brisken}, {Camilo},
  {Reynolds}, {van Kerkwijk}, {Halpern}, \& {Ransom}}]{hcb+07}
{Helfand}, D.~J., {Chatterjee}, S., {Brisken}, W.~F., {Camilo}, F., {Reynolds},
  J., {van Kerkwijk}, M.~H., {Halpern}, J.~P., \& {Ransom}, S.~M. 2007, \apj,
  662, 1198

\bibitem[{{Hobbs} {et~al.}(2005){Hobbs}, {Lorimer}, {Lyne}, \&
  {Kramer}}]{hllk05}
{Hobbs}, G., {Lorimer}, D.~R., {Lyne}, A.~G., \& {Kramer}, M. 2005, \mnras,
  360, 974

\bibitem[{{Hobbs} {et~al.}(2004){Hobbs}, {Lyne}, {Kramer}, {Martin}, \&
  {Jordan}}]{hlk+04}
{Hobbs}, G., {Lyne}, A.~G., {Kramer}, M., {Martin}, C.~E., \& {Jordan}, C.
  2004, \mnras, 353, 1311

\bibitem[{{Hoogerwerf} {et~al.}(2000){Hoogerwerf}, {de Bruijne}, \& {de
  Zeeuw}}]{hbz00}
{Hoogerwerf}, R., {de Bruijne}, J.~H.~J., \& {de Zeeuw}, P.~T. 2000, \apjl,
  544, L133

\bibitem[{{Hoogerwerf} {et~al.}(2001){Hoogerwerf}, {de Bruijne}, \& {de
  Zeeuw}}]{hbz01}
---. 2001, \aap, 365, 49

\bibitem[{{Hotan} {et~al.}(2006){Hotan}, {Bailes}, \& {Ord}}]{hbo06}
{Hotan}, A.~W., {Bailes}, M., \& {Ord}, S.~M. 2006, \mnras, 369, 1502

\bibitem[{{Janka} \& {Mueller}(1996)}]{jm96}
{Janka}, H.-T., \& {Mueller}, E. 1996, \aap, 306, 167

\bibitem[{{Johnston} {et~al.}(2005){Johnston}, {Hobbs}, {Vigeland}, {Kramer},
  {Weisberg}, \& {Lyne}}]{jhv+05}
{Johnston}, S., {Hobbs}, G., {Vigeland}, S., {Kramer}, M., {Weisberg}, J.~M.,
  \& {Lyne}, A.~G. 2005, \mnras, 364, 1397

\bibitem[{{Kaplan} {et~al.}(2008{\natexlab{a}}){Kaplan}, {Chatterjee},
  {Gaensler}, \& {Anderson}}]{kcga08}
{Kaplan}, D.~L., {Chatterjee}, S., {Gaensler}, B.~M., \& {Anderson}, J.
  2008{\natexlab{a}}, \apj, 677, 1201

\bibitem[{{Kaplan} {et~al.}(2008{\natexlab{b}}){Kaplan}, {Chatterjee}, {Hales},
  {Gaensler}, \& {Slane}}]{kch+08}
{Kaplan}, D.~L., {Chatterjee}, S., {Hales}, C., {Gaensler}, B.~M., \& {Slane},
  P.~O. 2008{\natexlab{b}}, ArXiv e-prints, 0810.4184

\bibitem[{{Kaplan} {et~al.}(2002){Kaplan}, {van Kerkwijk}, \&
  {Anderson}}]{kvka02}
{Kaplan}, D.~L., {van Kerkwijk}, M.~H., \& {Anderson}, J. 2002, \apj, 571, 447

\bibitem[{{Kaplan} {et~al.}(2007){Kaplan}, {van Kerkwijk}, \&
  {Anderson}}]{kva07}
---. 2007, \apj, 660, 1428

\bibitem[{{Kharchenko} {et~al.}(2005){Kharchenko}, {Piskunov}, {R{\"o}ser},
  {Schilbach}, \& {Scholz}}]{kpr+05}
{Kharchenko}, N.~V., {Piskunov}, A.~E., {R{\"o}ser}, S., {Schilbach}, E., \&
  {Scholz}, R.-D. 2005, \aap, 438, 1163

\bibitem[{{Kramer} {et~al.}(2003){Kramer}, {Lyne}, {Hobbs}, {L{\"o}hmer},
  {Carr}, {Jordan}, \& {Wolszczan}}]{klh+03}
{Kramer}, M., {Lyne}, A.~G., {Hobbs}, G., {L{\"o}hmer}, O., {Carr}, P.,
  {Jordan}, C., \& {Wolszczan}, A. 2003, \apjl, 593, L31

\bibitem[{{Lai} {et~al.}(2001){Lai}, {Chernoff}, \& {Cordes}}]{lcc01}
{Lai}, D., {Chernoff}, D.~F., \& {Cordes}, J.~M. 2001, \apj, 549, 1111

\bibitem[{{Lattimer} \& {Prakash}(2004)}]{lp04}
{Lattimer}, J.~M., \& {Prakash}, M. 2004, Science, 304, 536

\bibitem[{{Ma} {et~al.}(1998){Ma}, {Arias}, {Eubanks}, {Fey}, {Gontier},
  {Jacobs}, {Sovers}, {Archinal}, \& {Charlot}}]{mae+98}
{Ma}, C., {Arias}, E.~F., {Eubanks}, T.~M., {Fey}, A.~L., {Gontier}, A.-M.,
  {Jacobs}, C.~S., {Sovers}, O.~J., {Archinal}, B.~A., \& {Charlot}, P. 1998,
  \aj, 116, 516

\bibitem[{{Migliazzo} {et~al.}(2002){Migliazzo}, {Gaensler}, {Backer},
  {Stappers}, {van der Swaluw}, \& {Strom}}]{mgb+02}
{Migliazzo}, J.~M., {Gaensler}, B.~M., {Backer}, D.~C., {Stappers}, B.~W., {van
  der Swaluw}, E., \& {Strom}, R.~G. 2002, \apjl, 567, L141

\bibitem[{{Ng} \& {Romani}(2007)}]{nr07}
{Ng}, C.-Y., \& {Romani}, R.~W. 2007, \apj, 660, 1357

\bibitem[{{Ng} {et~al.}(2007){Ng}, {Romani}, {Brisken}, {Chatterjee}, \&
  {Kramer}}]{nrb+07}
{Ng}, C.-Y., {Romani}, R.~W., {Brisken}, W.~F., {Chatterjee}, S., \& {Kramer},
  M. 2007, \apj, 654, 487

\bibitem[{{Petrov} {et~al.}(2005){Petrov}, {Kovalev}, {Fomalont}, \&
  {Gordon}}]{vcs3}
{Petrov}, L., {Kovalev}, Y.~Y., {Fomalont}, E., \& {Gordon}, D. 2005, \aj, 129,
  1163

\bibitem[{{Preston} {et~al.}(1983){Preston}, {Morabito}, \& {Jauncey}}]{pmj83}
{Preston}, R.~A., {Morabito}, D.~D., \& {Jauncey}, D.~L. 1983, \apj, 269, 387

\bibitem[{{Romney}(1999)}]{r99}
{Romney}, J.~D. 1999, in Astronomical Society of the Pacific Conference Series,
  Vol. 180, Synthesis Imaging in Radio Astronomy II, ed. G.~B. {Taylor}, C.~L.
  {Carilli}, \& R.~A. {Perley}, 57--+

\bibitem[{{Socrates} {et~al.}(2005){Socrates}, {Blaes}, {Hungerford}, \&
  {Fryer}}]{sbhf05}
{Socrates}, A., {Blaes}, O., {Hungerford}, A., \& {Fryer}, C.~L. 2005, \apj,
  632, 531

\bibitem[{{Spangler} {et~al.}(2002){Spangler}, {Kavars}, {Kortenkamp}, {Bondi},
  {Mantovani}, \& {Alef}}]{skk+02}
{Spangler}, S.~R., {Kavars}, D.~W., {Kortenkamp}, P.~S., {Bondi}, M.,
  {Mantovani}, F., \& {Alef}, W. 2002, \aap, 384, 654

\bibitem[{{Splaver} {et~al.}(2005){Splaver}, {Nice}, {Stairs}, {Lommen}, \&
  {Backer}}]{sns+05}
{Splaver}, E.~M., {Nice}, D.~J., {Stairs}, I.~H., {Lommen}, A.~N., \& {Backer},
  D.~C. 2005, \apj, 620, 405

\bibitem[{{Standish}(1982)}]{s82}
{Standish}, Jr., E.~M. 1982, \aap, 114, 297

\bibitem[{{Taylor} \& {Cordes}(1993)}]{tc93}
{Taylor}, J.~H., \& {Cordes}, J.~M. 1993, \apj, 411, 674

\bibitem[{{Thorsett} {et~al.}(2002){Thorsett}, {Brisken}, \& {Goss}}]{tbg02}
{Thorsett}, S.~E., {Brisken}, W.~F., \& {Goss}, W.~M. 2002, \apjl, 573, L111

\bibitem[{{Toscano} {et~al.}(1999){Toscano}, {Britton}, {Manchester}, {Bailes},
  {Sandhu}, {Kulkarni}, \& {Anderson}}]{tbm+99}
{Toscano}, M., {Britton}, M.~C., {Manchester}, R.~N., {Bailes}, M., {Sandhu},
  J.~S., {Kulkarni}, S.~R., \& {Anderson}, S.~B. 1999, \apjl, 523, L171

\bibitem[{{Verbiest} {et~al.}(2008){Verbiest}, {Bailes}, {van Straten},
  {Hobbs}, {Edwards}, {Manchester}, {Bhat}, {Sarkissian}, {Jacoby}, \&
  {Kulkarni}}]{vbv+08}
{Verbiest}, J.~P.~W., {Bailes}, M., {van Straten}, W., {Hobbs}, G.~B.,
  {Edwards}, R.~T., {Manchester}, R.~N., {Bhat}, N.~D.~R., {Sarkissian}, J.~M.,
  {Jacoby}, B.~A., \& {Kulkarni}, S.~R. 2008, \apj, 679, 675

\bibitem[{{Vlemmings} {et~al.}(2004){Vlemmings}, {Cordes}, \&
  {Chatterjee}}]{vcc04}
{Vlemmings}, W.~H.~T., {Cordes}, J.~M., \& {Chatterjee}, S. 2004, \apj, 610,
  402

\bibitem[{{Wall} {et~al.}(2005){Wall}, {Jackson}, {Shaver}, {Hook}, \&
  {Kellermann}}]{wjs+05}
{Wall}, J.~V., {Jackson}, C.~A., {Shaver}, P.~A., {Hook}, I.~M., \&
  {Kellermann}, K.~I. 2005, \aap, 434, 133

\bibitem[{{Walter} \& {Lattimer}(2002)}]{wl02}
{Walter}, F.~M., \& {Lattimer}, J.~M. 2002, \apjl, 576, L145

\bibitem[{{Winkler} \& {Petre}(2007)}]{wp07}
{Winkler}, P.~F., \& {Petre}, R. 2007, \apj, 670, 635

\bibitem[{{Yakovlev} \& {Pethick}(2004)}]{yp04}
{Yakovlev}, D.~G., \& {Pethick}, C.~J. 2004, \araa, 42, 169

\bibitem[{{Zeiger} {et~al.}(2008){Zeiger}, {Brisken}, {Chatterjee}, \&
  {Goss}}]{zbcg08}
{Zeiger}, B.~R., {Brisken}, W.~F., {Chatterjee}, S., \& {Goss}, W.~M. 2008,
  \apj, 674, 271

\end{thebibliography}
\end{document}